\documentclass[preprint,aps,showpacs,pre,amsmath,amssymb]{revtex4}

\usepackage{amssymb}
\usepackage{graphicx}
\usepackage{dsfont}
\usepackage{color,soul}

\begin{document}
	
	\title{Electronic properties of twisted multilayer graphene}
	
	\author{V. Hung Nguyen$^1$, Trinh X. Hoang$^2$ and J.-C. Charlier$^1$}
	\affiliation{$^1$Institute of Condensed Matter and Nanosciences, Universit\'{e} catholique de Louvain (UCLouvain), B-1348 Louvain-la-Neuve, Belgium \\
		$^2$Institute of Physics, Vietnam Academy of Science and Technology, Ba Dinh, Hanoi 11108, Vietnam}
	
	\begin{abstract}
    Twisted bilayer graphene displays many fascinating properties that can be tuned by varying the relative angle (also called twist angle) between its monolayers. As a remarkable feature, both the electronic flat bands and the corresponding strong electron localization have been obtained at a specific ``magic'' angle ($\sim 1.1^{\circ}$), leading to the observation of several strongly correlated electronic phenomena. Such a discovery has hence inspired the creation of a novel research field called \textit{twistronics}, i.e., aiming to explore novel physical properties in vertically stacked  2D structures when tuning the twist angle between the related layers. In this paper, a comprehensive and systematic study related to the electronic properties of twisted multilayer graphene (TMG) is presented based on atomistic calculations. The dependence of both the global and the local electronic quantities on the twist angle and on the stacking configuration are analyzed, fully taking into account atomic reconstruction effects. Consequently, the correlation between structural and electronic properties are clarified, thereby highlighting the shared characteristics and differences between various TMG systems as well as providing a comprehensive and essential overview. On the basis of these investigations, possibilities to tune the electronic properties are discussed, allowing for further developments in the field of \textit{twistronics}.
	\end{abstract}

	\maketitle
	
	\section{Introduction}
	
	Twisted multilayer graphene (TMG) refers to structures of vertically stacked graphene layers when one or a few layers are rotated relative to the others by arbitrary angles \cite{wu2014,chen2016,mogera2017,shen2020,liu2020,park2021,zhang2021,chen2021e,zeyu2021,maria2021,jeong2021}, thus creating the so-called moiré superlattices. The first member of TMG systems, twisted bilayer graphene (TBLG), has recently attracted great interest from the scientific community since it exhibits fascinating properties, leading to the exploration of several novel features as surprising as superconductivity that can not be observed in conventional graphene systems \cite{andrei2020}. In particular and most remarkably, electronic properties of TBLG can be tuned by varying the twist angle $\theta$ \cite{wang2019,carr2020}, especially leading to the observation of localized flat bands near the Fermi level at a critical angle ($\simeq 1.1^\circ$) called \textit{the magic angle} \cite{rafi2011,utama2021}. At this critical angle, many-body effects are therefore significantly enlarged \cite{choi2018,kerelsky2019,choi2021,gadelha2021,gadelha2021b,barbosa2022} and charge carriers do not possess enough kinetic energy to escape from their strong mutual interactions, thus forming novel strongly correlated electronic states. Such a strong electronic correlation in magic-angle TBLG is the key ingredient for observing several novel phenomena such as superconductivity, correlated insulating states, magnetism, and even quantized anomalous Hall effects \cite{cao2018a,cao2018b,yankowitz2019,xie2019,lu2019,choi2019,jiang2019,aaron2019,serlin2020,leon2020}. 

    The observations mentioned above have been considered as real scientific breakthroughs, creating a nascent field of \textit{twistronics} \cite{carr2017}. Indeed, after the discovery of magic-angle TBLG, the research of other twisted systems has become an emerging topic. In addition to TMG, the explored systems include also twisted superlattices of other 2D layered materials such as graphene on hexagonal boron nitride, graphene on transition-metal dichalcogenide layers, van der Waals moir\'e superlattices of transition-metal dichalcogenide layers, and so on (e.g., see Refs.~\cite{liu2014,kang2017,naik2018,xian2019,yaping2020,kennes2021,xu2021,tao2022} as well as the recent review \cite{liu2022} and references therein). These explorations have also inspired researchers to extendedly apply the twist approach to other physical systems, e.g., magnetic 2D materials \cite{tong2018,wang2020,tiancheng2021,luo2021} and photonic layered structures \cite{wang2020p,mao2021,zhang2021p,chen2021}. In overall, the properties of those systems are definitely dependent of their constituting layers and importantly, can be tuned by varying their twist angle, similar to those observed in TBLG. In particular, in the TMG cases, electronic flat bands and accordingly, the  strongly correlated electronic phenomena have been similarly observed around their magic angle (e.g., see in Refs.~\cite{khalaf2019,haddadi2020,jeong2021,shen2020,youngju2020,chen2021e}). Moreover, more tunable possibilities can be observed in TMG \cite{bandwidth,yankowitz2019,liu2020,shen2020,chen2021e,zeyu2021,park2021,rickhaus2021}, due to a larger number of layers and the plenty of stacking configurations \cite{georgios2020}, compared to TBLG. In this context, a comprehensive understanding of the electronic properties of TMG is highly desirable. Some new TMGs \cite{zeyu2021,park2021,jeong2021} have been very recently produced experimentally, and hence their corresponding electronic properties are needed to be more exhaustively clarified. In addition, a good understanding of those twisted systems could be also helpful for research of twisted structures based on other 2D materials. 

    On the theoretical side, though a large number of works devoted to the electronic properties of some TMG has been reported (e.g., in \cite{carr2017,carr2019,khalaf2019,carr2020,tritsaris2020,haddadi2020,shin2021}), a systematical study on this subject is however still missing. Remarkably, it has been shown that atomic reconstruction (i.e., the structure relaxation) significantly occurs in TBLG at low angles ($\lesssim 1.1^\circ$), thus strongly influencing their electronic properties, i.e., especially the formation of electronic flat bands as well as the electronic localization pictures \cite{gargiulo2017,gadelha2021,nguyen2021}. The structural and electronic properties of TBLG also exhibit a strong correlation \cite{nguyen2021}, that explains essentially the corresponding electronic localization features. Fully taking into account the atomic reconstruction effects is therefore mandatory to compute accurately the electronic properties of TMG, which has not been properly considered yet in all works previously published in the literature.
    
    This paper aims to investigate systematically the most common and essential electronic properties of TMG by atomistic calculations, which fully take the atomic reconstruction into account. First, twisted bilayer graphene is investigated to illustrate the overall structural and electronic properties induced by the twisting effects. Some of these properties are commonly observed in other multilayer structures. Afterward, the study focuses on several typical twisted multilayer (more than 2 graphene layers) systems around their magic angles, at which electronic flat bands and strong electron localization are also observed. The correlation between the obtained electronic and structural properties is also clarified. Electronic (both global and local) quantities obtained depending on the twist angle and the stacking configuration are analyzed in details, thus highlighting the shared characteristics and differences between TMGs. Finally, on the basis of these investigations, possibilities to tune these electronic properties are discussed.
	
	\section{Modeling Methodology}
	
	\begin{figure*}[!b]
		\centering
		\includegraphics[width = 0.8\textwidth]{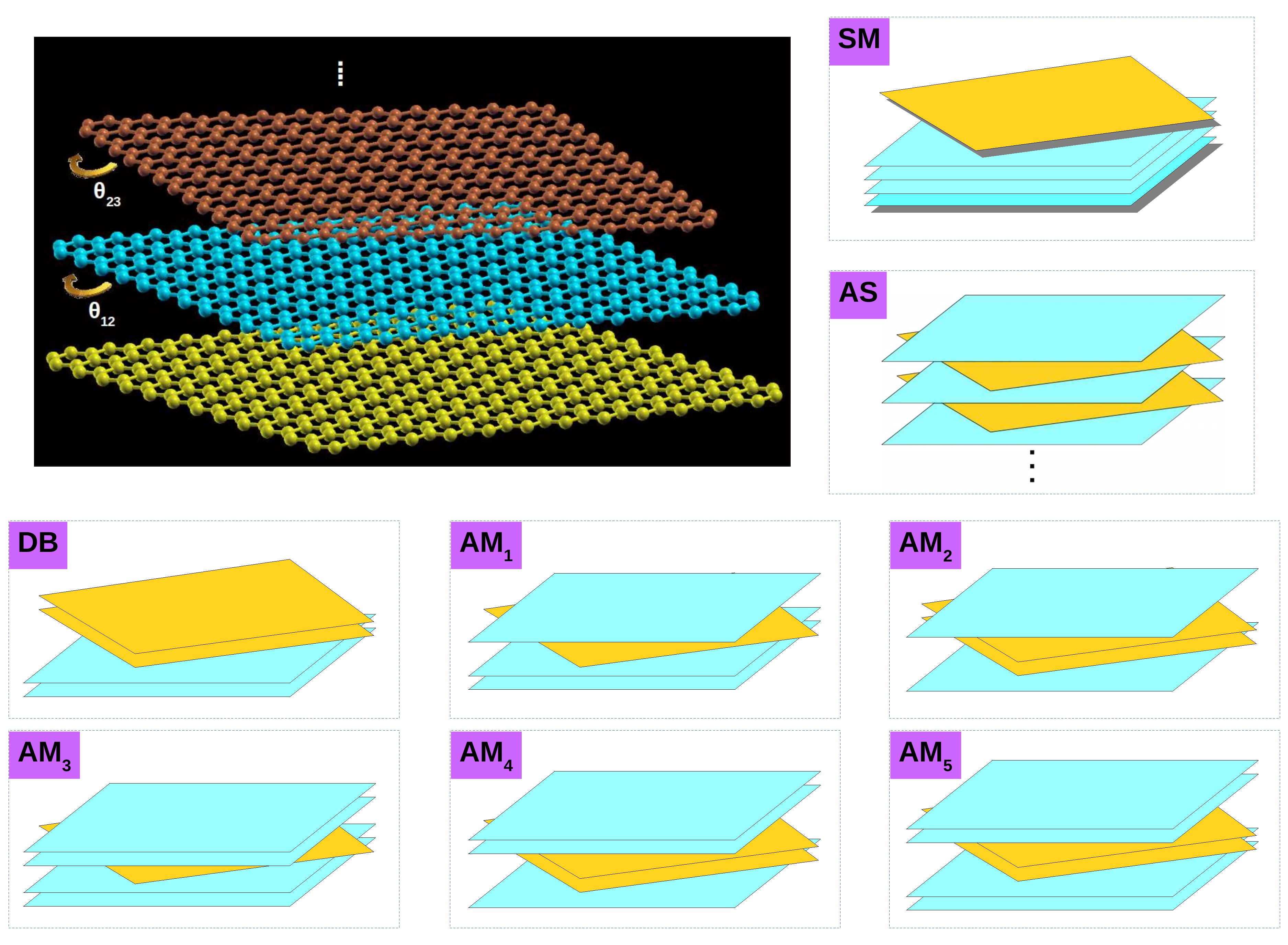}
		\caption{Twisted multilayer graphene systems investigated in this work.
		SM-structures are twisted systems created by putting and rotating a graphene monolayer on top of a multilayer (\textit{n} layers) graphene that is either in AB stacked or in ABC stacked. These structures are hereafter abbreviated as SM-(AB)$_n$ and SM-(ABC)$_n$, respectively. AS-structures are alternatively twisted systems of \textit{n} graphene monolayers, abbreviated as AS$_n$. DB is the abbreviation of twisted double bilayer graphene superlattices. At last, AM$_n$ structures (\textit{n} = 1,2,3,4,5) are alternatively twisted systems containing, at least, one Bernal stacking graphene bilayer.} 
		\label{fig_sim0}
	\end{figure*}
	As mentioned earlier, the atomic reconstruction is significant and hence taking into account these structural effects is mandatory to model accurately the electronic properties of twisted graphene systems, especially, at small twist angles around and below the magic one. In this work, we used classical force-fields to compute the atomic structure relaxation. More specifically, intralayer forces are computed using the optimized Tersoff and Brenner potentials \cite{lindsay2010}, while interlayer forces are modeled using the Kolmogorov-Crespi potentials \cite{kolmogorov2005,leven2016}. The atomic structure was optimized until all force components are smaller than 0.5 meV/atom and the electronic calculations were then performed. The atomic structure relaxation obtained by this method is shown to be in good agreement with experiments \cite{brihuega2012,kerelsky2019,yoo2019,zhang2020} as well as the results obtained by other (both quantum and semi-classical) methods \cite{uchida2014,gargiulo2017,cantele2020,lamparski2020}, except that the intralayer equilibrium \textit{C-C} bond distance is slightly overestimated by $\sim 1\%$ \cite{leven2016}.
	
	Similarly to previous works~\cite{gadelha2021,nguyen2021}, the electronic properties of TMG (schematized in Fig.\ref{fig_sim0}) were then computed using the $p_z$ tight-binding (TB) Hamiltonians after the structural optimization has been fully achieved. Hopping energies $t_{nm}$ between carbon \textit{C-}sites are determined by the standard Slater-Koster formula
	\begin{equation}
		t_{nm} = \cos^2 \phi_{nm} V_{pp\sigma} (r_{nm}) + \sin^2 \phi_{nm} V_{pp\pi} (r_{nm})
	\end{equation}
	where the direction cosine of $\vec r_{nm} = \vec r_{m} - \vec r_{n}$ along Oz axis is $\cos\phi_{nm} = z_{nm}/r_{nm}$. The distance-dependent Slater-Koster parameters are determined as \cite{trambly12}
	\begin{equation*}
		\begin{aligned}
			& V_{pp\pi} (r_{nm}) = V_{pp\pi}^0 \exp \left[ q_\pi \left( 1 - \frac{r_{nm}}{a_0}  \right) \right] F_c (r_{nm})  \\
			& V_{pp\sigma} (r_{nm}) = V_{pp\sigma}^0 \exp \left[ q_\sigma \left( 1 - \frac{r_{nm}}{d_0}  \right) \right] F_c (r_{nm})
		\end{aligned}
	\end{equation*}
	with a smooth cutoff function $F_c (r_{nm}) = \left[ 1 + \exp \left( \frac{r_{nm}-r_c}{\lambda_c} \right) \right]^{-1}$.
	
	In this work, the TB Hamiltonian was further simplified from the model considered in \cite{nguyen2021}, i.e., only couplings limited by $r_{inplane} \leq 2.16$ \AA $ $ were taken into account. This approximation presents an advantage that the matrix of TB Hamiltonian is very sparse so as to make calculations feasible for small twist angles and/or multilayer systems in which the number of \textit{C} atoms could be huge. Within this approximation, the TB parameters \cite{nguyen2021} are slightly adjusted as follows:
	\begin{equation*}
		\begin{aligned}
			& V_{pp\pi}^0 = -2.7 \,~\mathrm{eV}, \,\,\,  V_{pp\sigma}^0 = 367.5 \,~\mathrm{meV}, \\
			& \frac{q_\pi}{a_0} = \frac{q_\sigma}{d_0} = 22.18 \,~\mathrm{nm}^{-1}, \\ 
			& a_0 = 0.1439 \,~\mathrm{nm}, \,\,\, d_0 = 0.33 \,~\mathrm{nm}, \,\,\, r_c = 0.614 \,~\mathrm{nm}, \,\,\, \lambda_c = 0.0265 \,~\mathrm{nm}
		\end{aligned}
	\end{equation*}
	Indeed, such an adjusted TB Hamiltonian fairly accurately reproduces the electronic properties of TBLG as obtained in \cite{nguyen2021} as well as in experiments, e.g., in \cite{cao2018a,gadelha2021,barbosa2022}. As a clear illustration, the almost flat bands near the Fermi level is perfectly reproduced in the relaxed TBLG in the vicinity of the magic angle (i.e., $\sim 1.1^\circ$, see Fig. \ref{fig_sim2} below) as experimentally observed (e.g., see in Ref.~\cite{cao2018a}). 
	
	\section{Twisted bilayer graphene}
	
	\begin{figure*}[!ht]
		\centering
		\includegraphics[width = 0.8\textwidth]{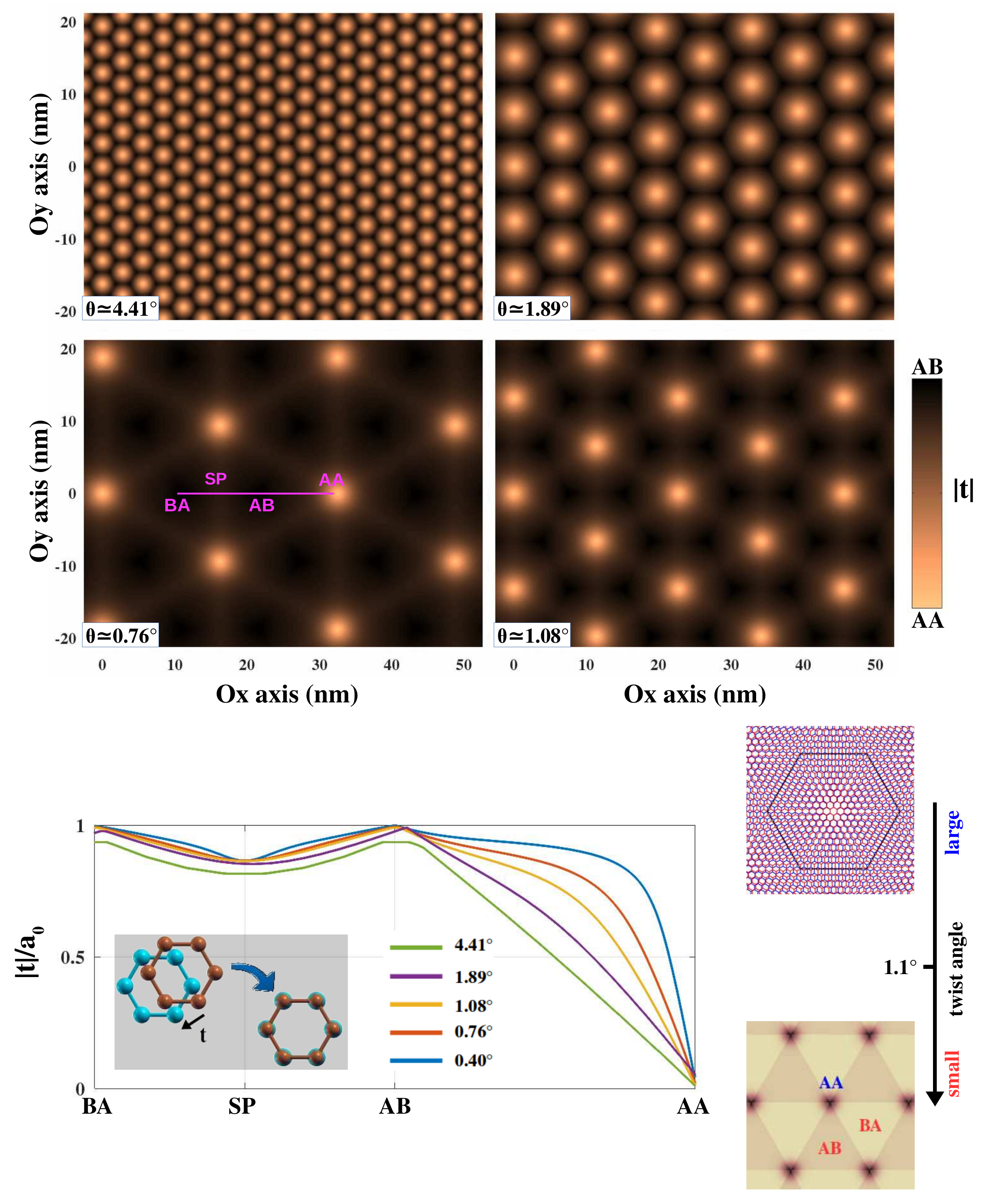}
		\caption{Stacking structure of TBLG imaged by the spatial variation of stacking vector \textbf{\textit{t}} (as defined in the inset of bottom-left panel) when varying the twist angle $\theta$. Four top images are 2D real-space representation of $|$\textbf{\textit{t}}$|$ while the bottom-left one is a plot of $|$\textbf{\textit{t}}$|$ extracted along the line connecting BA - SP (soliton line) - AB - AA stacking regions. The bottom-right diagram is a simple description of the stacking properties of TBLG in the different $\theta$-regimes (from large to small twist-angles).}
		\label{fig_sim1}
	\end{figure*}
	As the simplest TMG structure, TBLG exhibits most clearly essential electronic properties induced by the twisting effects. These properties as well as their correlation with the structural properties, that generally also appear in other TMGs (as examples, see in Refs.~\cite{nguyen2021,haddadi2020} as well as illustration in Figs.\ref{fig_sim13}-\ref{fig_sim14} in section V), will be therefore revised and systematized in this section.
	
	Typical images illustrating the evolution of the stacking structure in TBLG when varying the twist angle $\theta$ are presented in Fig.\ref{fig_sim1}. The stacking vector \textbf{\textit{t}}, as illustrated in the bottom-left panel, indicates the smallest displacement applied locally to one layer to recover the AA stacking at the considered position. In general, TBLG superlattice includes three typical stacking regions: AA stacking, AB/BA stacking regions and soliton lines between AB/BA ones (called SP for $saddle$ $point$). The presented images illustrate that the magic angle $\sim 1.1^\circ$ separates TBLG into two regimes: $\theta > 1.1^\circ$ (large angles) and $\theta < 1.1^\circ$ (small ones) \cite{nguyen2021}. At large angles, the moir\'e superlattice evolves smoothly and both AA and AB/BA stacking regions are continuously enlarged when decreasing the twist angle. However, AA stacking regions get saturation when $\theta$ decreasingly approaches $1.1^\circ$. At small angles, whereas the size of AA stacking regions is unchanged, AB/BA regions are continuously enlarged when decreasing $\theta$, according to the enlargement of moir\'e cells. Thus, in addition to the critical angle (i.e., magic angle) $\sim 1.1^\circ$, there are two TBLG classes, separated by such angle and exhibiting distinct stacking structures, as summarized in the bottom-right diagram of Fig.\ref{fig_sim1}.
	
	\begin{figure*}[!ht]
		\centering
		\includegraphics[width = 0.98\textwidth]{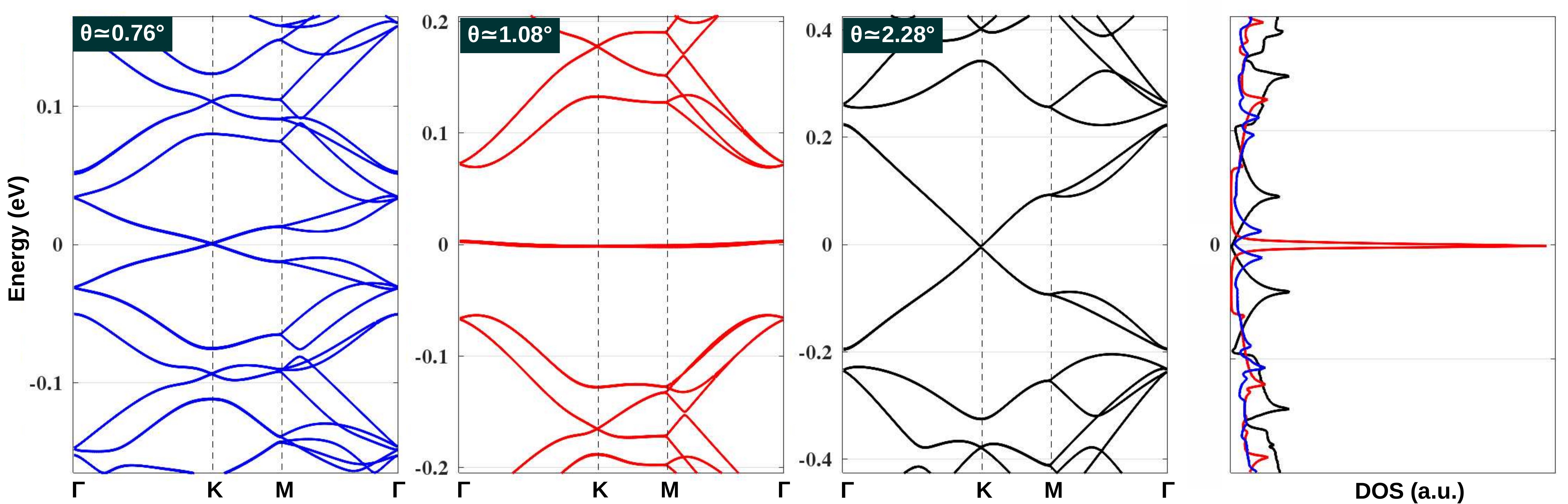}
		\caption{Electronic bandstructure of TBLG for the twist angle $\theta < 1.1^\circ$ (left in blue), $\theta \simeq 1.1^\circ$ (center in red) and $\theta > 1.1^\circ$ (right in black). The corresponding density of states (DOS) are presented on the extreme right with the same color scale for the three moir\'e structures.}
		\label{fig_sim2}
	\end{figure*}
\begin{figure*}[!ht]
		\centering
		\includegraphics[width = 0.8\textwidth]{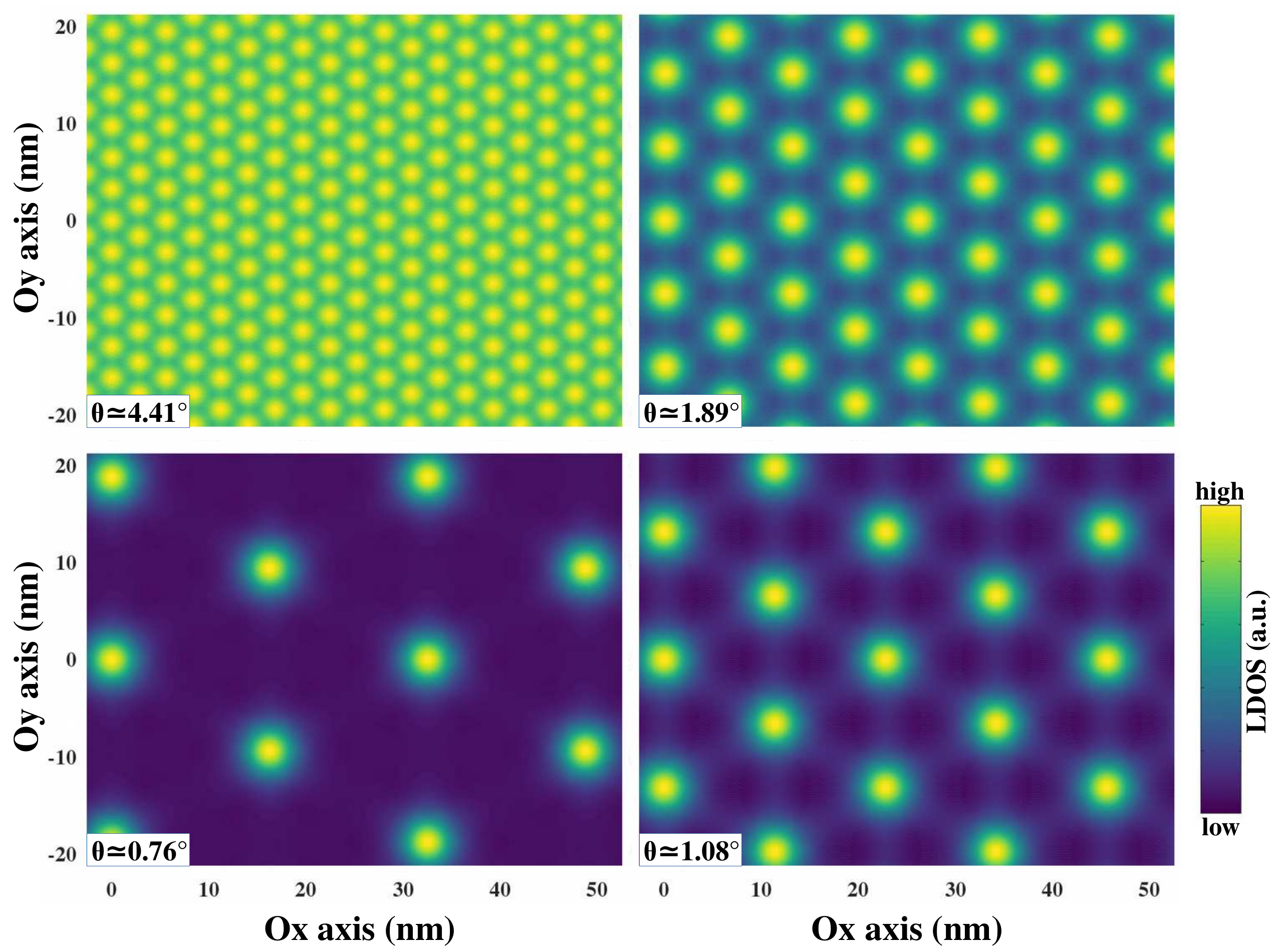}
		\caption{Real-space LDOS images in TBLG at different twist angles. Energy is fixed where the strongest electronic localization is obtained, i.e., at the van Hove singularity point in DOS (see Fig.\ref{fig_sim2}) for $\theta > 1.1^\circ$ and at zero energy for $\theta \lesssim 1.1^\circ$.}
		\label{fig_sim3}
	\end{figure*}

	The structural properties discussed above crucially govern the electronic features of TBLG, as illustrated by the typical electronic structures in Fig.\ref{fig_sim2} as well as the local densities of states (LDOS) images in Fig.\ref{fig_sim3}. In large-angle TBLG, Dirac fermions are still preserved at low energies but a remarkable renormalization of their Fermi velocity is obtained \cite{yin2015}, as seen in Fig.\ref{fig_sim2} for $\theta \simeq 2.28^\circ$. In addition, saddle points emerge at the crossing of Dirac cones (i.e., at the M-point in the superlattice Brillouin region), yielding van Hove singularities in the density of states (DOS) at energies that are lowered down when decreasing $\theta$. When $\theta$ approaches $1.1^\circ$, two (electron and hole) van Hove singularities merge at zero energy and accordingly the low energy bands are flattened, resulting in the observation of an extremely high DOS peak at that energy (see Fig.\ref{fig_sim2} for $\theta \simeq 1.08^\circ$). The angle $1.1^\circ$ has been thus called the ``magic angle'' \cite{rafi2011}. When further decreasing $\theta$, the electronic bands become more dispersive again (see Fig.\ref{fig_sim2} for $\theta \simeq 0.76^\circ$). Remarkably, it has been shown that fully taking the structure relaxation into account, the electronic flat band (accordingly, the localized DOS peak at zero energy) as at $1.1^\circ$ is no longer observed for small angles $< 1.1^\circ$ as discussed in details in Ref.~\cite{nguyen2021}. Moreover, even though it is not present in the global electronic quantities, the strong electronic localization is still observed locally in AA stacking regions, which has been experimentally confirmed in Ref.~\cite{gadelha2021}.
	
	These distinct electronic properties of TBLG at large and small angles are visually illustrated by the LDOS pictures in Fig.\ref{fig_sim3} as well as their correlation with the structural properties discussed in Fig.\ref{fig_sim1}. Actually, the LDOS in Fig.\ref{fig_sim3} are represented in each case at the strongest localization point. First, the strong electronic localization is observed in the AA stacking region for all twist angles. Remarkably, the real-space electronic localization exhibits the similar $\theta$-dependence, compared to that of stacking structure presented in Fig.\ref{fig_sim1}. Indeed, while it is continuously enlarged when reducing $\theta$ above $1.1^\circ$, the real-space LDOS peak obtained in AA stacking regions is maximum at $\theta \simeq 1.1^\circ$ and saturates in the regime $< 1.1^\circ$. These results demonstrate clearly a strong correlation between the structural properties and the electronic localization features in TBLG. 
	
	More remarkably, the observation of electronic flat bands is found to be essentially related to the maximum electronic localization in AA stacking regions and their maximum contribution to the global electronic properties of the system (see also discussions in section V). These features are concurrently obtained at $\theta \simeq 1.1^\circ$. To clarify such observation, we notice additionally here that the strongest electronic localization in energy is favorable (unfavorable) in AA stacking regions (other stacking regions) \cite{nguyen2021}. At large angles $> 1.1^\circ$, the electronic localization in AA stacking regions (see Fig.\ref{fig_sim3} for $\theta \simeq 4.41^\circ$) and consequently the global localization in energy (see the extreme right panel of Fig.\ref{fig_sim2}) are relatively weak. When decreasing $\theta$ in this regime, the electronic localization in AA stacking regions and their contribution are continuously enhanced. Accordingly, global localization (i.e., DOS) peaks increase and their positions concurrently move towards zero energy. At $\theta \simeq 1.1^\circ$, all these effects are maximized and consequently the electronic flat bands are obtained. When further decreasing $\theta$ in the regime below $1.1^\circ$, even though a strong electronic localization is still locally observed in the saturated AA stacking region, the increasing contribution of AB/BA stacking ones gradually diminishes the global electronic localization and hence flat bands can not be observed again \cite{nguyen2021}, thus precluding the emergence of other (i.e., small) magic angles predicted by other theoretical calculations \cite{rafi2011}.

	Finally, it is worth noting that even though they could be quantitatively affected by the presence of a larger number of layers and/or by a more complicatedly stacked configuration, the structural and electronic features similar to those presented above are still observed in other TMGs, simply because all these features are essentially governed by twisting effects.
	
	\section{Twisted multilayer graphene}
	
    In this section, the electronic properties of TMGs schematized in Fig.\ref{fig_sim0} are investigated. The creation of these systems could be simply considered as adding one or a few graphene layers to a TBLG one. Since these TMG share a similar $\theta$-dependence (e.g., see Refs.~\cite{haddadi2020,nguyen2021}) with TBLG, the present study is mainly focused on the magic-angle cases where low-energy bands are most flat and thus strongly correlated electronic phenomena can be observed.
    \begin{figure*}[!ht]
		\centering
		\includegraphics[width = 0.98\textwidth]{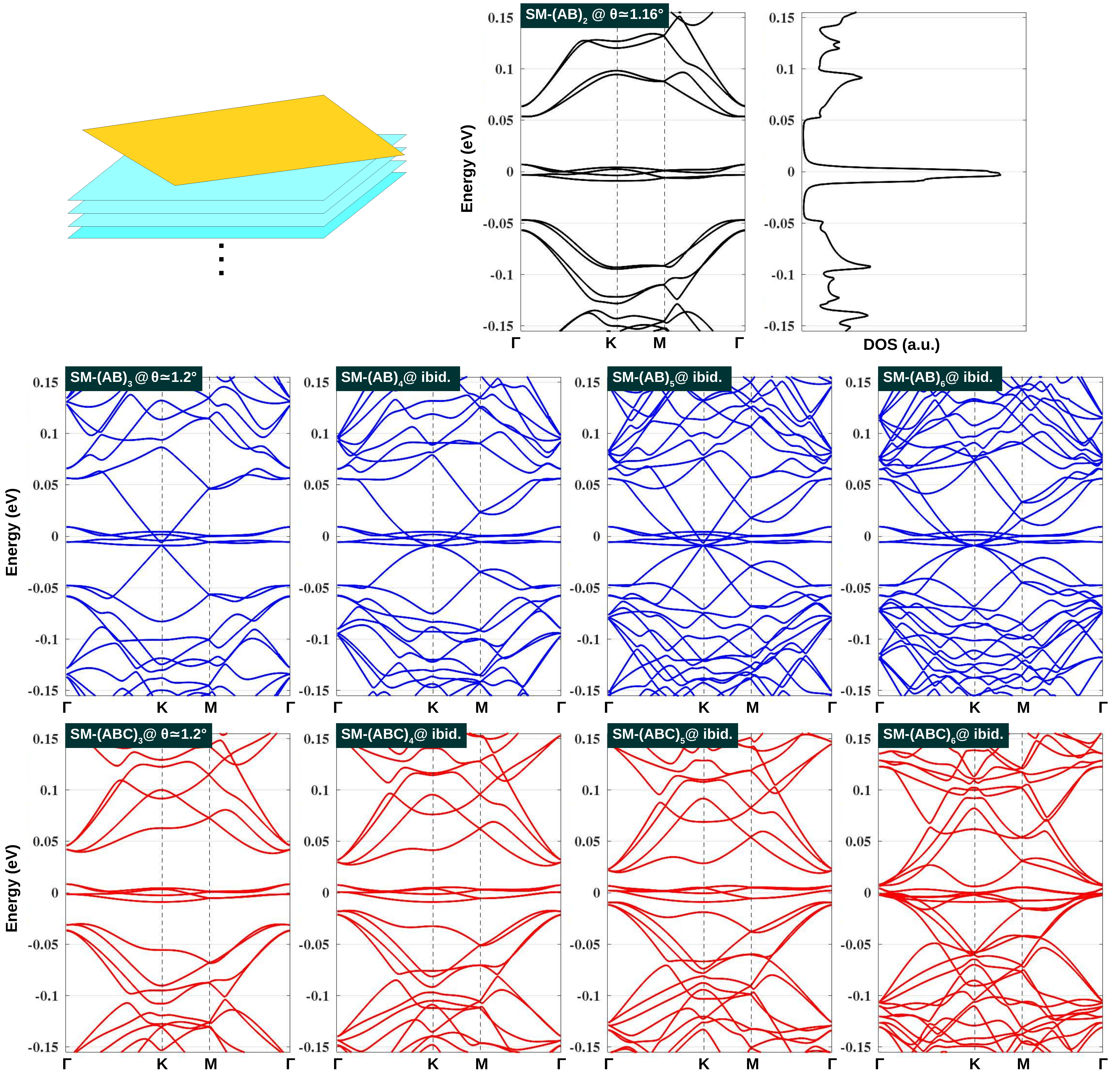}
		\caption{Electronic bandstructures of different SM-systems at their magic angle. On top, the results in SM-(AB)$_2$ (i.e., twisted monolayer-bilayer) structure are presented. Other systems including SM-(AB)$_n$ and SM-(ABC)$_n$ (see the definition in Fig.\ref{fig_sim0}) at their magic angle $\sim 1.2^\circ$ are computed and presented in the second and third rows, respectively, with \textit{n} = 3,4,5,6 (from left to right). On top-right, DOS versus energy in SM-(AB)$_2$ structure is also superimposed.}
		\label{fig_sim4}
	\end{figure*}

	We first consider TMG obtained when one rotated graphene monolayer is placed on top of a graphene multilayer, which is either in an AB-stacked form or an ABC-stacked one (i.e., SM-systems in Fig.\ref{fig_sim0}). Fig.\ref{fig_sim4} presents the electronic bandstructures of different SM-systems in both two cases. First of all, the SM-(AB)$_2$ (i.e., twisted monolayer-bilayer) system at its magic angle exhibits a bandstructure (see on top of Fig.\ref{fig_sim4}) with isolated flat bands, similar to that obtained in the magic-angle TBLG. The main difference between two cases is that these bands (accordingly, zero-energy DOS peak) are observed to be less flat (lower) than those of TBLG. Note that continuum Hamiltonian calculations \cite{youngju2020} have shown that the bandwidth of flat bands in SM-(AB)$_2$ system is generally narrower than that obtained in magic-angle TBLG. This is indeed also confirmed by our calculations for systems at the same twist angle in the regime $\theta >> 1.1^\circ$. The results in the magic angle case however exhibit an opposite behavior, i.e., the bandwidth of the magic-angle
	\begin{figure*}[!ht]
		\centering
		\includegraphics[width = 0.8\textwidth]{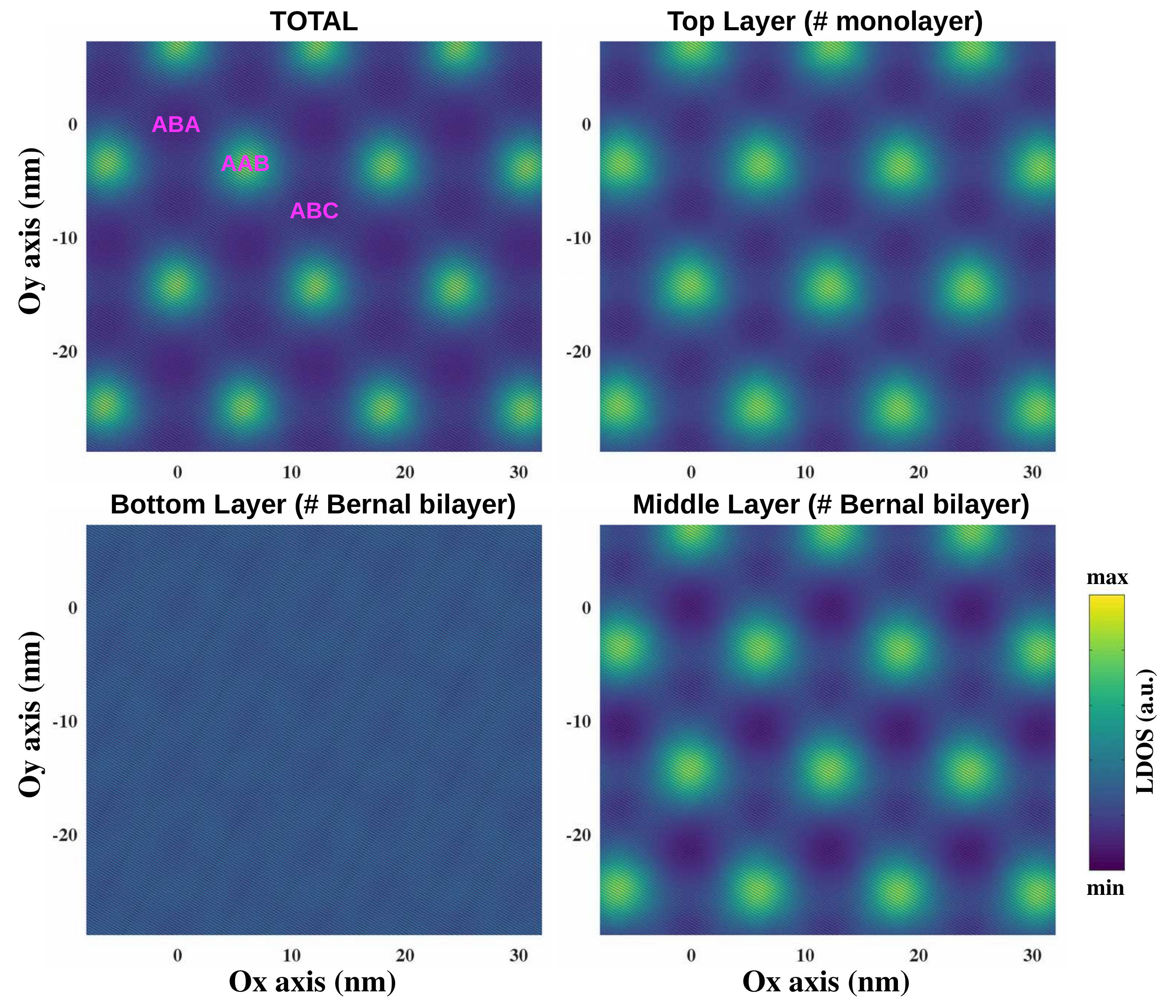}
		\caption{Total LDOS and layer decomposed LDOS in SM-(AB)$_2$ system in Fig.\ref{fig_sim4} at zero energy. }
		\label{fig_sim5}
	\end{figure*}
    SM-(AB)$_2$ system is larger than that obtained in magic-angle TBLG. This is actually consistent with the results presented in Refs.~\cite{choi2021,bandwidth} for atomic reconstructed superlattices. These results could be understood as a consequence of the fact, as discussed in the previous section, that the band flatness has a strong correlation with the real-space electronic localization that is basically stronger in AA-stacking regions \cite{nguyen2021}. In SM-(AB)$_2$ systems, the corresponding regions are obtained in an AAB-stacked form, i.e., a mixture of AA- and AB-stacking configurations while the latter is unfavorable for the electronic localization. Such mixture diminishes the electronic localization in the SM-(AB)$_2$ system. This is visibly illustrated by comparing the total LDOS in Fig.\ref{fig_sim5} with that obtained in magic-angle TBLG (see Fig.\ref{fig_sim3} for $\theta \simeq 1.08^\circ$) and is further demonstrated in Fig.\ref{fig_sim13} in section V, thus explaining essentially the less flat bands obtained in the SM-(AB)$_2$ case.
	
	In addition, another interesting feature is found in Fig.\ref{fig_sim5} that flat bands in the magic-angle SM-(AB)$_2$ system exhibit a coexistence of layer-resolved localized-delocalized behaviors, i.e., they are spatially localized on the twisted side whereas delocalized on the bilayer side. This unique feature has been also experimentally confirmed in a recent work \cite{tong2021} and the delocalized correlated electronic states have been similarly explored in twisted double bilayer graphene \cite{zhang2021b}. Note that electronic states delocalized either in the graphene plane or in different layers are generally obtained in twisted systems consisting of more than 2 layers as the SM-(AB)$_2$ system here and others considered below, where certain layers are strongly influenced by the moir\'e superlattice interactions whereas the weak effects take place in other layers. Moreover, the observed coexistence of localized-delocalized electronic states essentially has a tight relationship with the weak real-space localization in the SM-(AB)$_2$ system discussed above.
	
	\begin{figure*}[!ht]
		\centering
		\includegraphics[width = 0.98\textwidth]{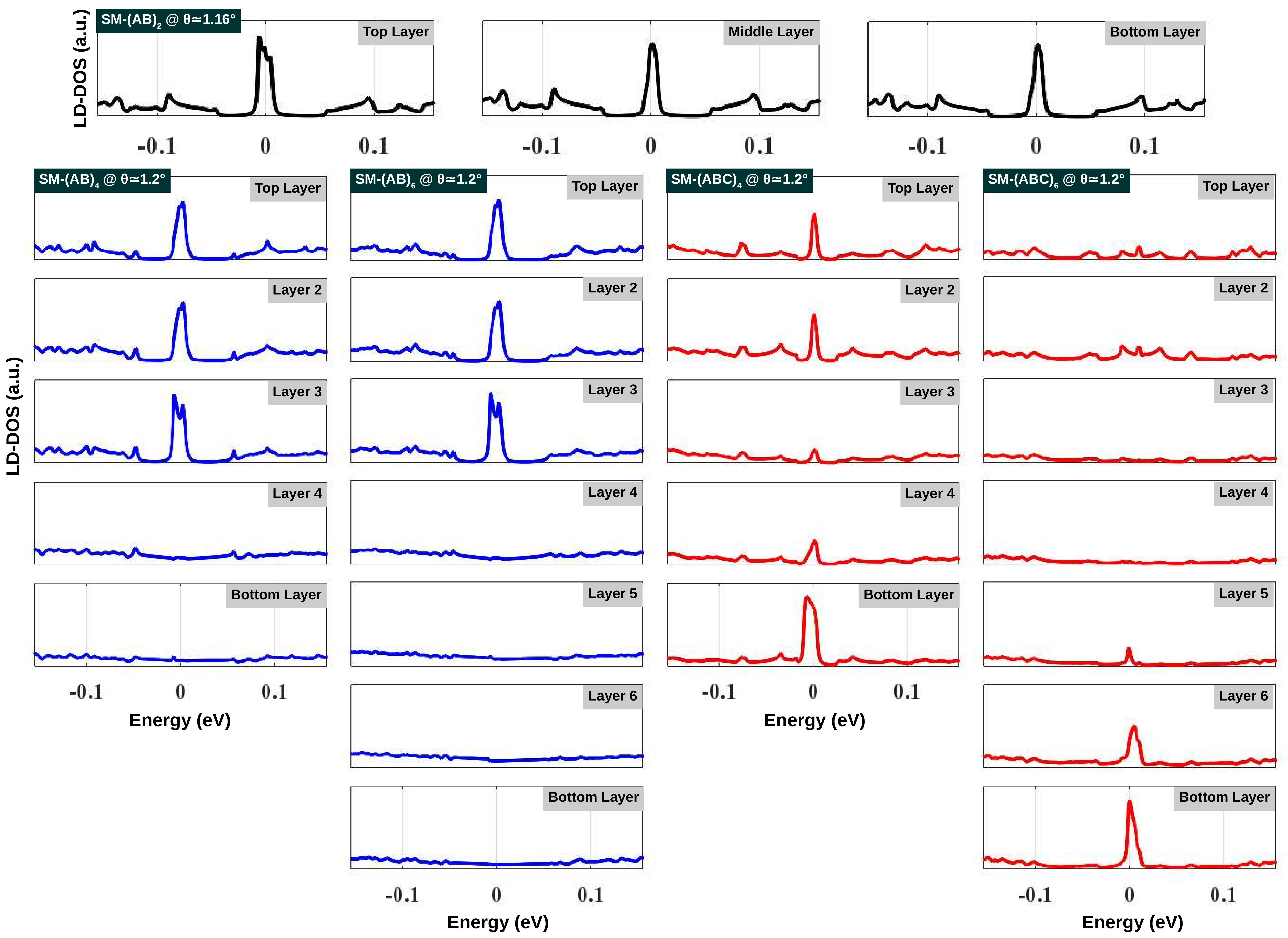}
		\caption{Layer decomposed DOS (LD-DOS) in some SM-systems considered in Fig.\ref{fig_sim4}. }
		\label{fig_sim6}
	\end{figure*}
	When the number of graphene layers of the multilayer counterpart increases, both systems SM-(AB)$_n$ and SM-(ABC)$_n$ exhibit clearly distinct electronic properties (see Fig.\ref{fig_sim4}). In particular, while the electronic structure of SM-(AB)$_n$ systems presents both flat bands and dispersive ones at low energies, flat bands in SM-(AB)$_n$ systems are still clearly isolated from high-energy bands for $n \leq 5$. However, the energy gaps between the mentioned bands in the latter case is reduced when increasing \textit{n} and is closed for $n > 5$. Such distinct properties could be explained by the difference in electronic properties of AB- and ABC-stacked graphene multilayer systems, i.e., while ABC-stacked systems host a single band that is partially flat around zero energy, there is a mixture of several (increasing with \textit{n}) parabolic and linear bands in the AB-stacked ones \cite{koshino2010}.

	To further clarify the difference between these two systems, we analyze their layer decomposed DOS (LD-DOS) in function of energy in Fig.\ref{fig_sim6}. For the SM-(AB)$_2$ system, besides the coexistence of layer-resolved localized-delocalized states presented in Fig.\ref{fig_sim5}, LD-DOS still exhibits the usual behaviours similar to that observed in magic-angle TBLG. However, other distinct features are found for SM-systems with a larger number of graphene layers. In particular, the electron localization is mostly observed in three top layers of SM-(AB)$_n$ structures where the moir\'e superlattice interactions take place. However, the localization is obtained in both top and bottom outermost layers of SM-(ABC)$_3$ and SM-(ABC)$_4$ systems (see illustration in Fig.\ref{fig_sim6} for SM-(ABC)$_4$ one). Moreover, when the number of layers \textit{n} further increases, such electron localization in SM-(ABC)$_n$ surprisingly changes, i.e., the effect in the top outermost layers is diminished while it is enhanced in the bottom ones. When $n$ is large enough, the electron localization is mostly observed in the bottom outermost layers as illustrated in Fig.\ref{fig_sim6} for the SM-(ABC)$_6$ system.
	\begin{figure*}[!ht]
		\centering
		\includegraphics[width = 0.98\textwidth]{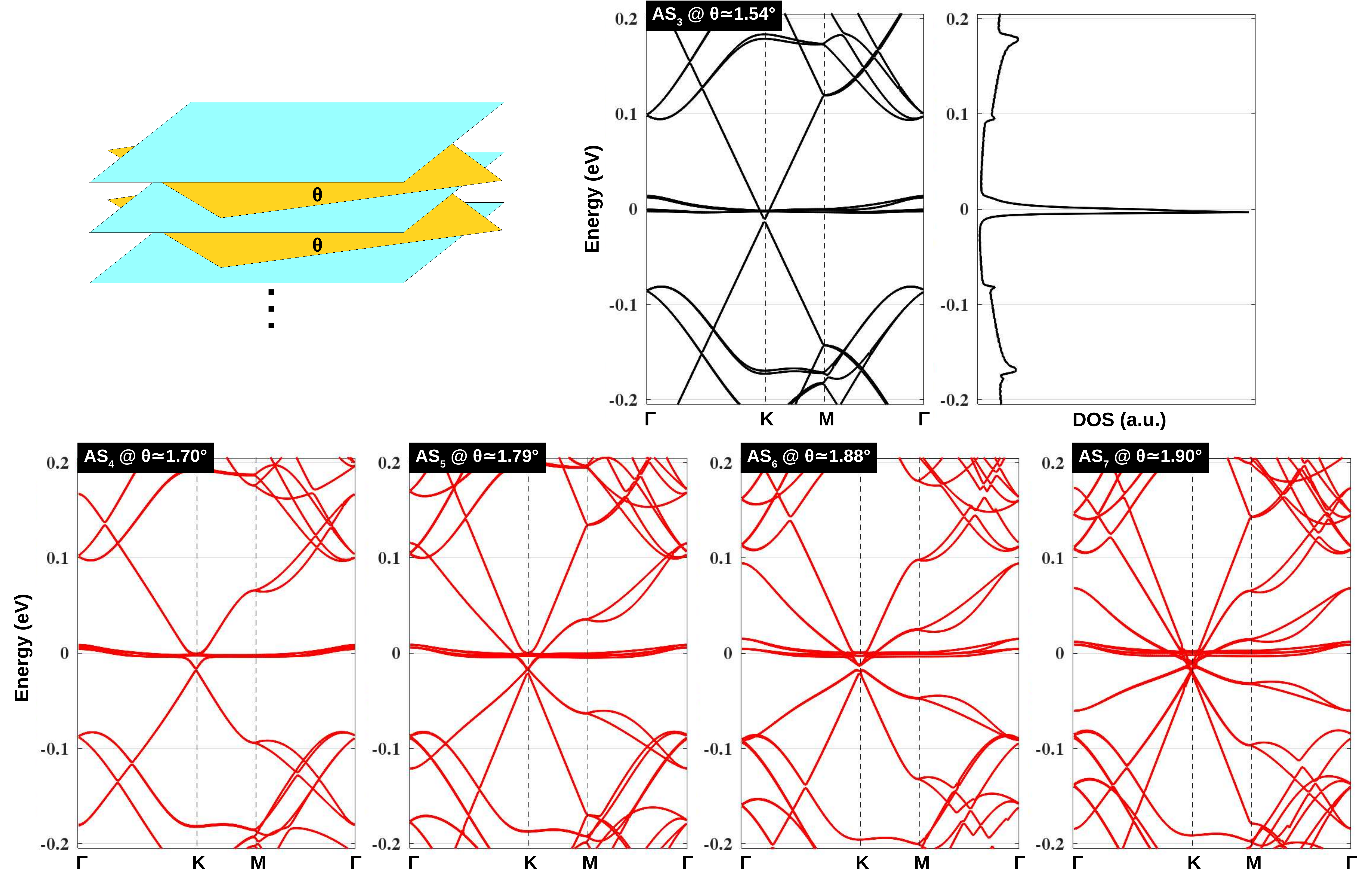}
		\caption{Electronic bandstructures of alternatively twisted systems of graphene monolayers schematized on top-left (i.e., the AS$_n$ systems specified in Fig.\ref{fig_sim0}) with $\theta$ near their magic angle $\theta_n^{MA}$ (see text). On top-right, DOS versus energy in AS$_2$-structure is also superimposed.}
		\label{fig_sim7}
	\end{figure*}
	
	Next, TMGs created by alternatively rotating ($0,\theta,0,\theta,...$) graphene monolayers (i.e., AS$_n$ structures in Fig.\ref{fig_sim0}) are considered in Fig.\ref{fig_sim7}. The magic angle of these AS$_n$ structures can be estimated by the simple expression $\theta_n^{MA} = 2 \theta_n^{MA} \cos (\pi / (n+1)))$ with $\theta_2^{MA} \simeq 1.1^\circ$ as proposed in Refs.~\cite{khalaf2019,jeong2021}. This is indeed demonstrated in Fig.\ref{fig_sim7} where (almost) flat bands are obtained in AS$_n$ structures near $\theta_n^{MA}$. Different from the results suggested by the effective continuum models \cite{khalaf2019,park2021,zeyu2021,jeong2021}, the TB bandstructure here presents four degenerate flat bands (more visibly seen in section V when a vertical electric field is applied) and \textit{n}-2 additional linear bands at low energies. Moreover, saddle points (accordingly, van Hove singularities in DOS) occur in dispersive bands for \textit{n}$\geqslant$4. The low-energy bands in these AS$_n$ systems are also shown to be generally flatter than those obtained for the SM-structures presented in Fig.\ref{fig_sim4}, thus confirming again our discussion above about the property of electron localization in the presence of Bernal stacking bilayer graphene. Note that another difference between systems without and with Bernal stacking bilayer graphene has been also discussed in Ref.~\cite{choi2021}. Indeed, in this study, significant sublattice polarization has been observed in the latter case, leading to weaker electron-phonon couplings than those obtained in twisted systems of graphene monolayers only.
	\begin{figure*}[!ht]
		\centering
		\includegraphics[width = 0.98\textwidth]{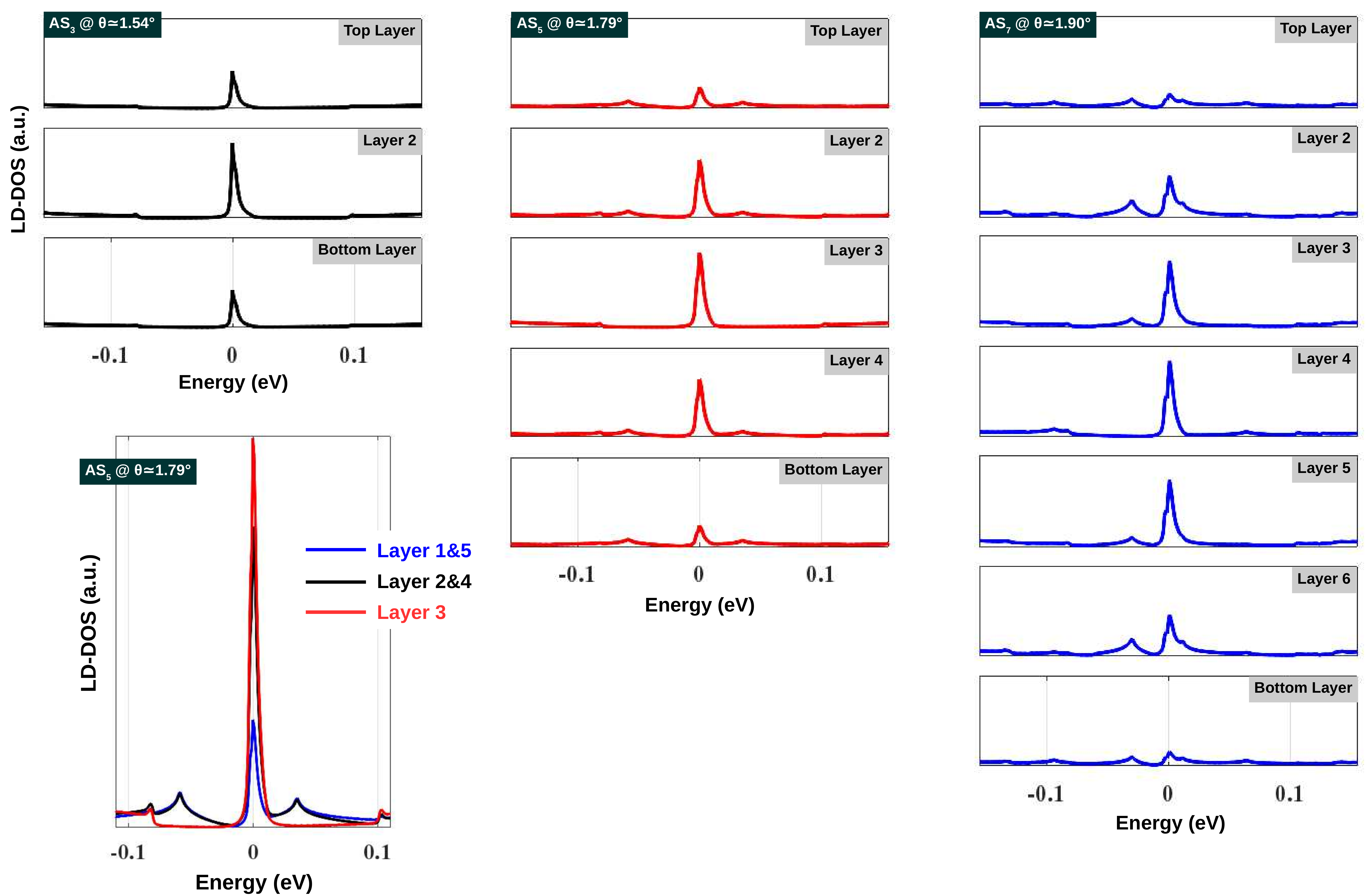}
		\caption{Layer decomposed DOS (LD-DOS) in some AS-systems considered in Fig.\ref{fig_sim7}. On bottom-left, a zoom-combined picture of the results in AS$_5$ structure is presented.}
		\label{fig_sim8}
	\end{figure*}
	
	To understand more deeply their electronic properties, the layer decomposed DOS obtained in these AS$_n$ structures is presented in Fig.\ref{fig_sim8}. Remarkably, the strong electronic localization takes place in their centered layers, in contrast to those observed in SM-ones presented in Fig.\ref{fig_sim6}. Moreover, LD-DOS in the centered layers presents behaviors that are very similar to those obtained in magic-angle TBLG. In particular, the strongly localized DOS peak is likely isolated from high-energy electronic states by finite energy gaps. Contrastingly, LD-DOS in other layers (especially, the outermost ones) exhibit some signatures of dispersive bands, i.e., the characteristics of linear bands and van Hove singularities induced by the twist as visibly shown in the zoom-image on the bottom of Fig.\ref{fig_sim8}. Thus, the flat bands and dispersive ones of AS$_n$ structures exhibit oppositely spatial properties, i.e., whereas the former is localized in the system center, the latter is observed in outer layers. This implies that there is aslo a coexistence of localized-delocalized electronic states in the AS$_n$ systems of large \textit{n}.
	\begin{figure*}[!ht]
		\centering
		\includegraphics[width = 0.78\textwidth]{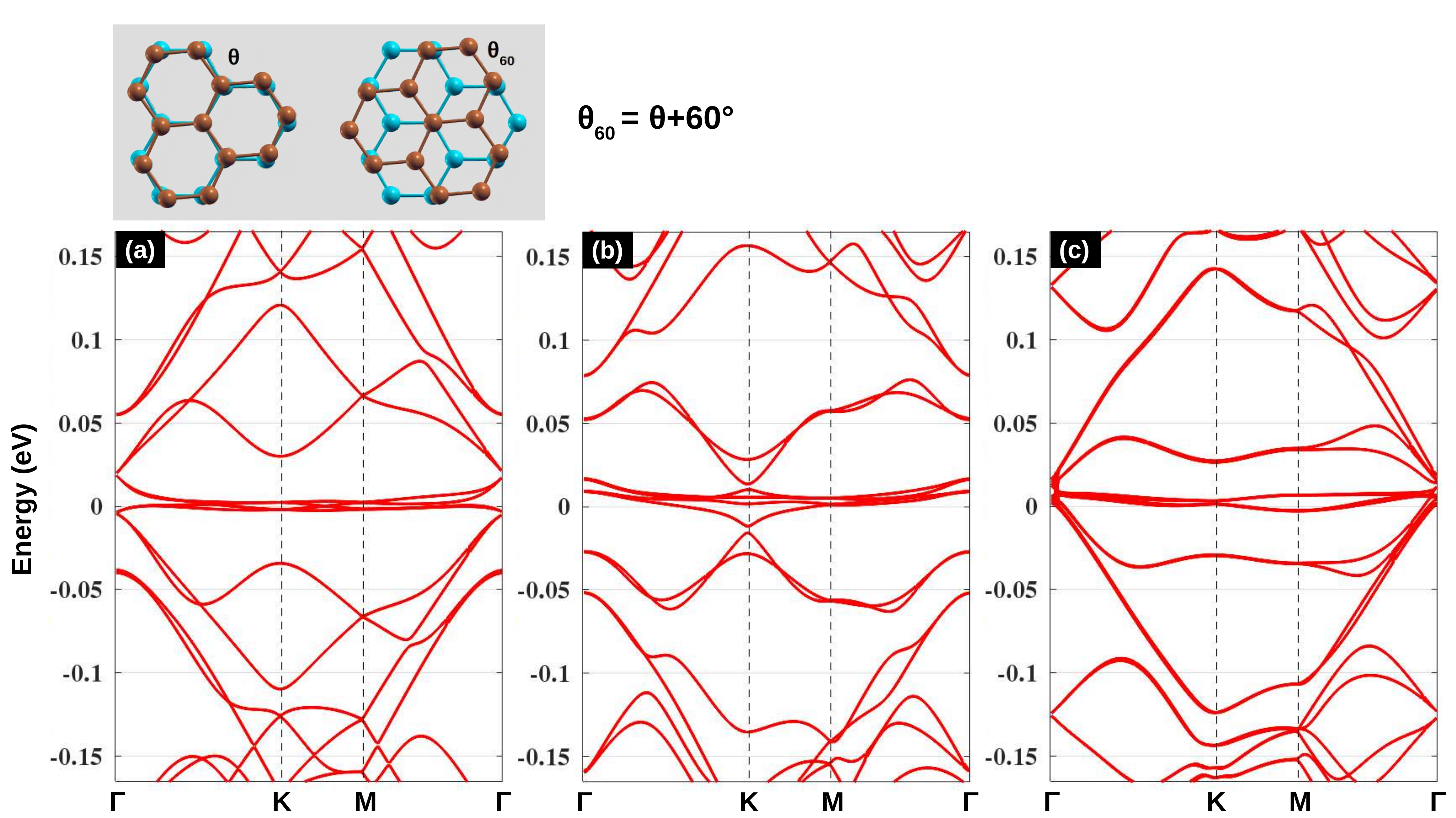}
		\caption{Electronic bandstructures of AS$_n$ systems when one or a few layers are additionally rotated by $60^\circ$ (see the illustration on top). In (a,b), the top layer of AS$_3$ at $\theta \simeq 1.30^\circ$ and AS$_4$ at $\theta \simeq 1.54^\circ$, respectively, is additionally rotated by $60^\circ$. In (c), such additional $60^\circ$-rotation is applied to two top layers of AS$_4$ at $\theta \simeq 1.35^\circ$.}
		\label{fig_sim9}
	\end{figure*}
	\begin{figure*}[!ht]
		\centering
		\includegraphics[width = 0.8\textwidth]{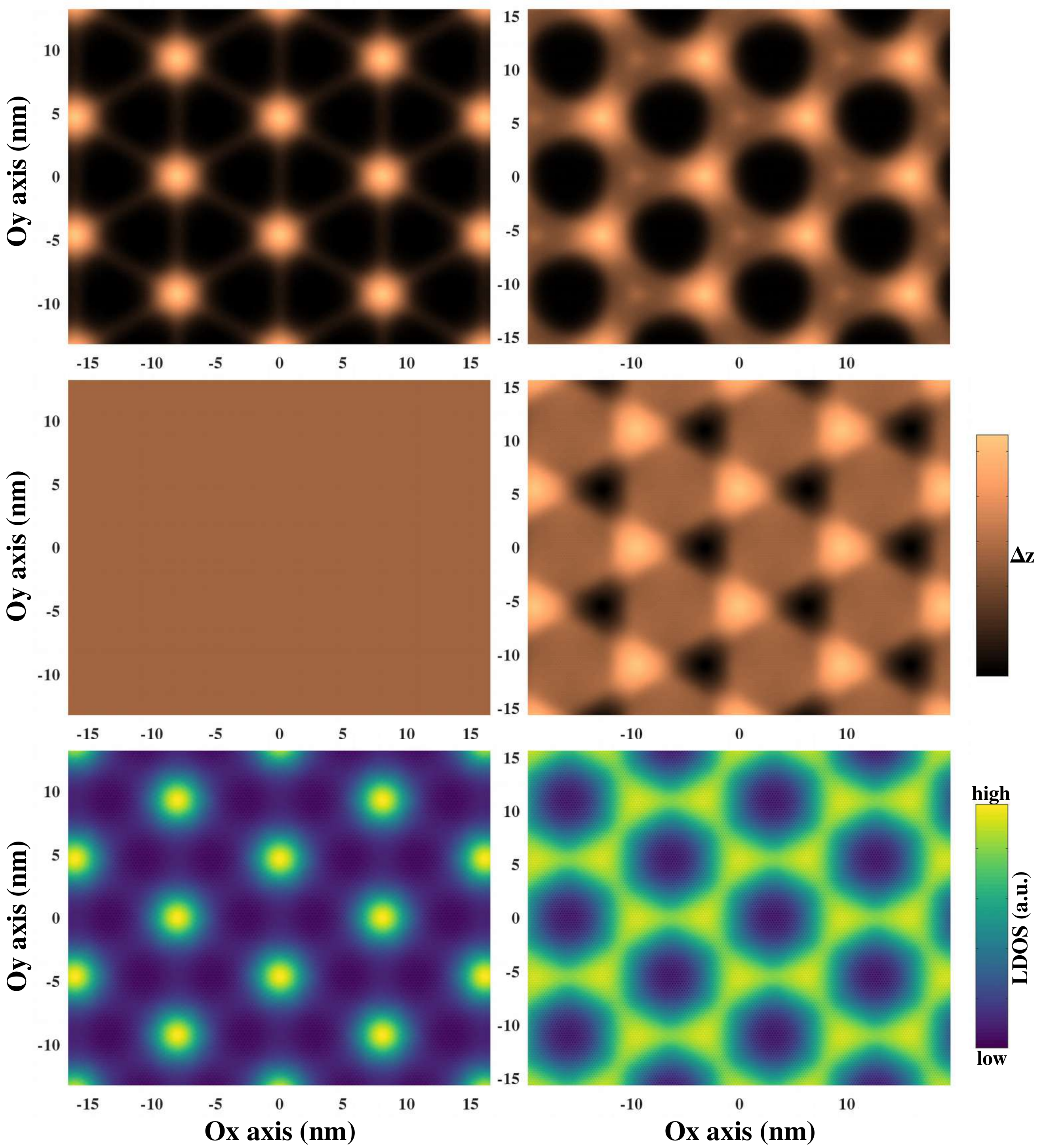}
		\caption{The structural and electronic properties of two AS$_3$ structures: AS$_3$ at $\theta \simeq 1.54^\circ$ in Fig.\ref{fig_sim7} (see in the left) and AS$_3$ at $\theta \simeq 1.30^\circ$ considered in Fig.\ref{fig_sim9}a (see in the right). Images in the first and second rows present the spatial variation of out-of-plane displacements in the top and middle layers, respectively. LDOS of two systems at zero energy is presented in the bottom images.}
		\label{fig_sim10}
	\end{figure*}
	
	When AS$_n$ systems are created, another remarkable possibility could in principle be obtained, in particular, one or a few graphene layers are rotated by $\theta_{60} = \theta$+60$^\circ$ (instead of $\theta$) as illustrated on top of Fig.\ref{fig_sim9}, similar to that reported in Ref.~\cite{finney2019} for graphene/hBN structures. Actually, such $\theta_{60}$-rotations do not change the lattice orientation of the graphene layer, compared to the simple $\theta$-rotated systems considered in Figs.\ref{fig_sim7}-\ref{fig_sim8}. However, three AAA-, ABA- and BAB-stacking regions obtained in the latter case are changed to AAB-, ABB- and ABC-stacking ones in the AS$_n$ systems with $\theta$+60$^\circ$ rotations, leading to significant changes in their electronic structure. Indeed, the flat electronic bands are observed at another twist angle as shown in Fig.\ref{fig_sim9}, compared to similar structures (i.e., with the same number of layers) presented in Figs.\ref{fig_sim7}-\ref{fig_sim8}. In addition, dispersive bands at low energies are generally no longer linear ones. These changes can be clarified by analyzing the structural properties of two considered types of AS$_n$ systems (e.g., see in Fig.\ref{fig_sim10} for the AS$_3$ cases). In particular, when compared to the first type (i.e., with $\theta$ rotations only), the top layer of the AS$_3$ system in Fig.\ref{fig_sim9}a presents a totally different buckling pattern (see the top panels in Fig.\ref{fig_sim10}). Similarly, the middle layer of the latter system is no longer flat as that observed in the former one (see the panels in the second row of Fig.\ref{fig_sim10}). Accordingly to these different structural properties, these two systems exhibit totally different LDOS images as shown in the bottom of Fig.\ref{fig_sim10}. These results thus clarify the difference in their electronic structures discussed above. 
	\begin{figure*}[!ht]
		\centering
		\includegraphics[width = 0.78\textwidth]{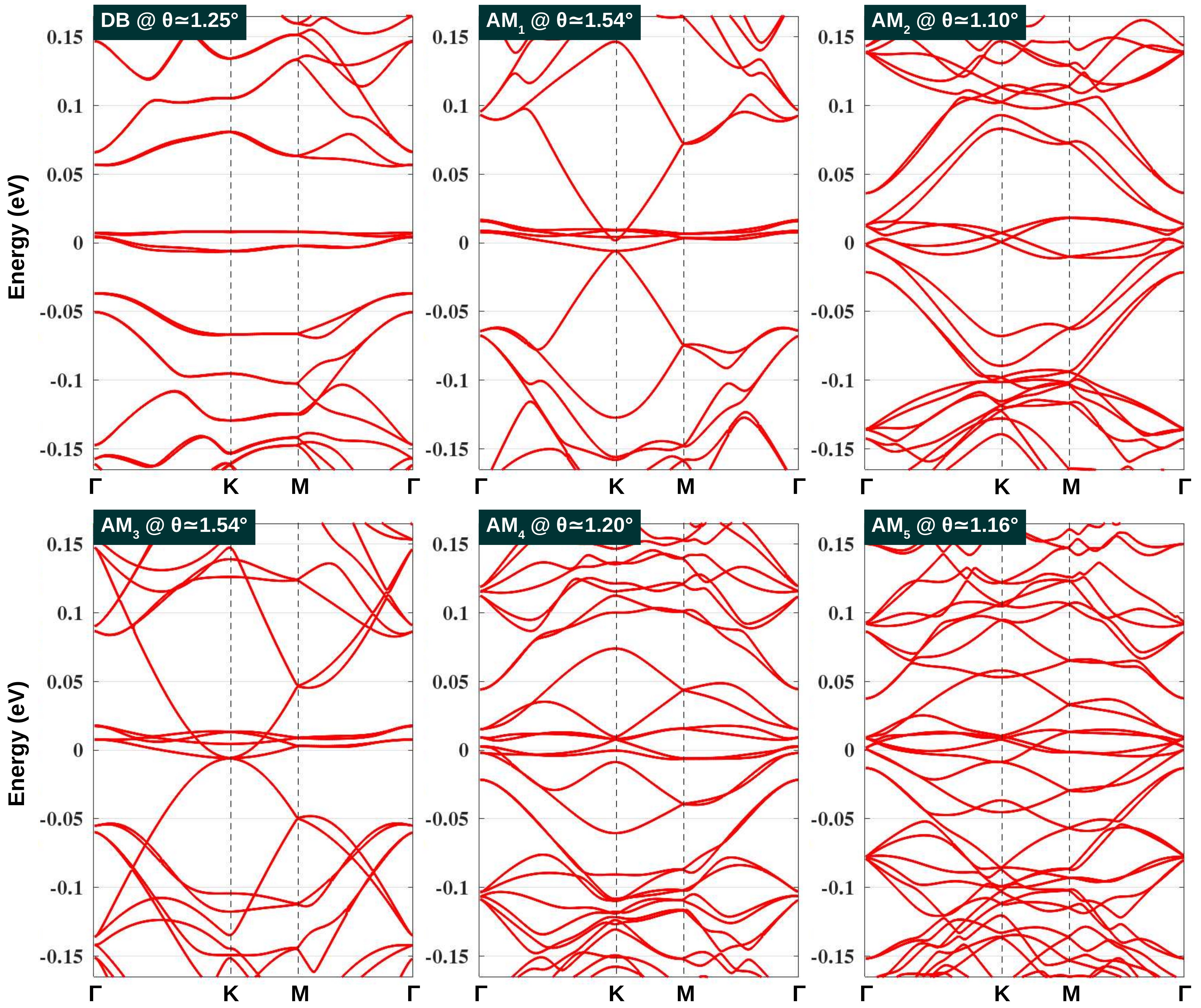}
		\caption{Electronic bandstructures of TMGs including DB- (i.e., twisted double bilayer graphene) and AM-structures (alternatively twisted structures containing, at least, one Bernal stacking bilayer, as schematized in Fig.\ref{fig_sim0}). Calculations were performed at their magic angles where low-energy bands are most flat.}
		\label{fig_sim11}
	\end{figure*}
	
	At last, the electronic bandstructures of other TMGs schematized in Fig.\ref{fig_sim0} at their magic angles are presented in Fig.\ref{fig_sim11}. First, the isolated flat bands are only obtained in magic-angle DB-structure (i.e., twisted double bilayer graphene) \cite{haddadi2020}, similar to those obtained in TBLG and in SM-(AB)$_2$ one. AM$_n$-systems (n=1,2,3,4,5) are actually alternatively twisted structures but different from the AS ones, containing at least one Bernal stacking bilayer. Most remarkably, the results obtained in those systems show that the flat bands can be easily obtained if the middle layer is a graphene monolayer (i.e., AM$_1$ and AM$_3$ structures), whereas the low-energy bands are generally less flat if it is a Bernal stacking bilayer. Moreover, similar to the SM-(AB)$_n$ and AS$_n$ structures, dispersive bands are always additionally obtained at low energies, which is likely an inherent property of alternatively twisted systems. Once more, the low-energy bands of those systems in the presence of Bernal stacking bilayers are shown to be less flat than those presented above for systems of graphene monolayers only.
	
	\begin{figure}[!ht]
		\centering
		\includegraphics[width = 0.86\textwidth]{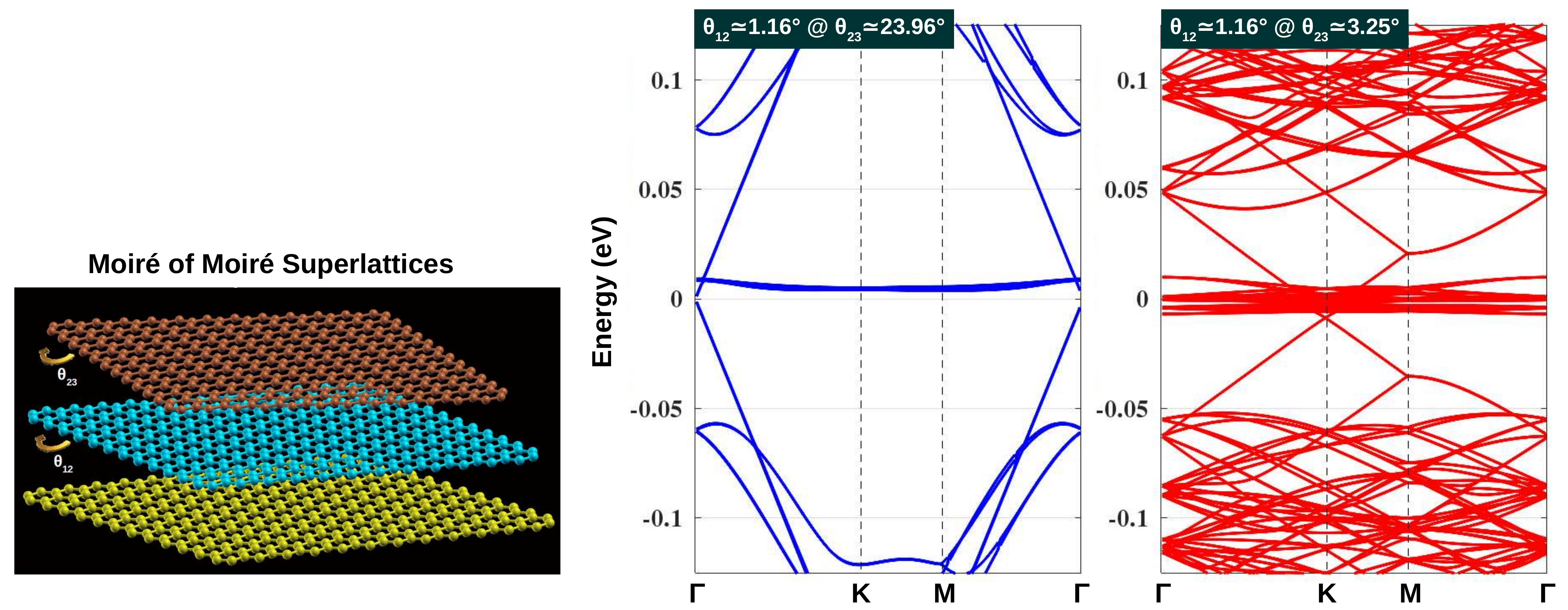}
		\caption{Electronic bandstructures of moir\'{e} of moir\'{e} graphene superlattices created by putting one rotated (by $\theta_{23}$) graphene monolayer on top of magic-angle TBLG (i.e., $\theta_{12} \simeq 1.08^\circ$). Two typical cases, particularly, large angle ($\theta_{23} \simeq 23.96^\circ$) and small angle ($\theta_{23} \simeq 3.25^\circ$), are presented.}
		\label{fig_sim12}
	\end{figure}
	For completeness, we finally present in Fig.\ref{fig_sim12} the electronic bandstructures of moir\'e-of-moir\'e graphene superlattices, which are here trilayer graphene systems created by vertically stacking magic-angle TBLG (i.e., $\theta_{12} \simeq 1.08^\circ$) and one rotated (by the angle $\theta_{23}$) graphene monolayer. We particularly consider two cases: a large angle $\theta_{23} \simeq 23.96^\circ$ and a small one $\theta_{23} \simeq 3.25^\circ$. Actually, when $\theta_{23}$ is large, the interaction between the graphene monolayer and the magic angle TBLG has no significant effect at low energies, i.e., the obtained low-energy bands are approximately a simple combination of the bands of those two systems. Actually, this feature is very similar to those obtained in TBLGs in the large angle regime when their bands can be considered as a combination of the bands of two graphene monolayers. However, when $\theta_{23}$ is small (i.e., $3.25^\circ$ considered here), the mentioned interaction has significant effects on the electronic structure, in particular, the bandwidth of flat bands is significantly enlarged, compared to that obtained in magic-angle TBLG. These results suggest that besides twisted graphene systems investigated above, other more complicated twisted multilayer ones, forming moir\'e-of-moir\'e superlattices, shall generally exhibit more complicated electronic structures with less opportunity for observing the flat bands and accordingly, strong electron localization.
	
	\section{Flattening electronic bands and tunability}
	
	As emphasized in section III, the electronic flat bands in reconstructed TBLG is essentially obtained in the situation when the real-space localization in AA stacking regions and their contribution to the global electronic properties are concurrently maximized at the magic angle. In fact, as it will be demonstrated below, such feature is commonly observed in TMGs and hence could be considered as the practical origin of their observed flat bands.

	\begin{figure}[!ht]
		\centering
		\includegraphics[width = 0.98\textwidth]{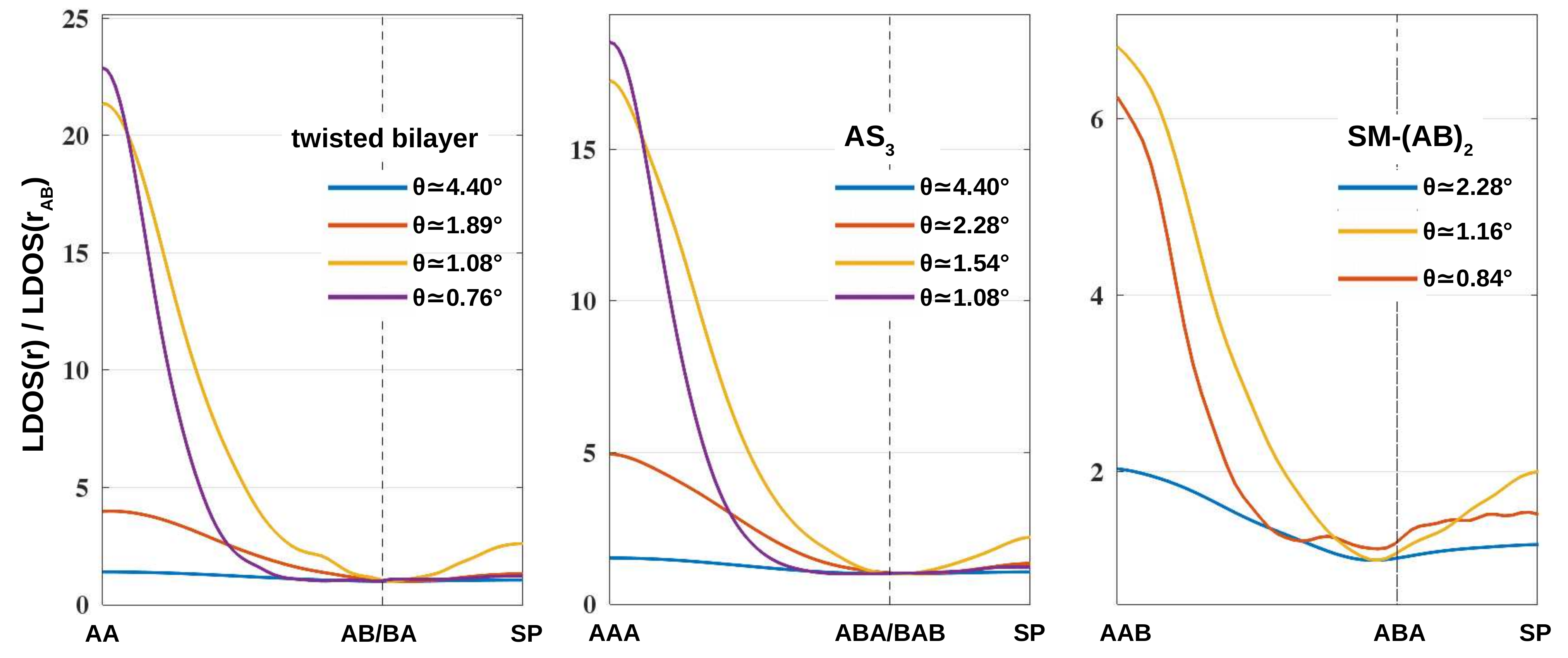}
		\caption{Spatial variation of LDOS in TBLG, AS$_3$ and SM-(AB)$_2$ systems when varying the twist angle around their magic one. The presented LDOS is extracted along the line connecting regions where AA, SP, AB/BA stacking configurations occur. Calculations are performed at energy where the strongest electronic localization is obtained, i.e., at the electron van Hove singularity point for large $\theta$ and at zero energy for $\theta$ below the magic angle.}
		\label{fig_sim13}
	\end{figure}
	\begin{figure}[!b]
		\centering
		\includegraphics[width = 0.98\textwidth]{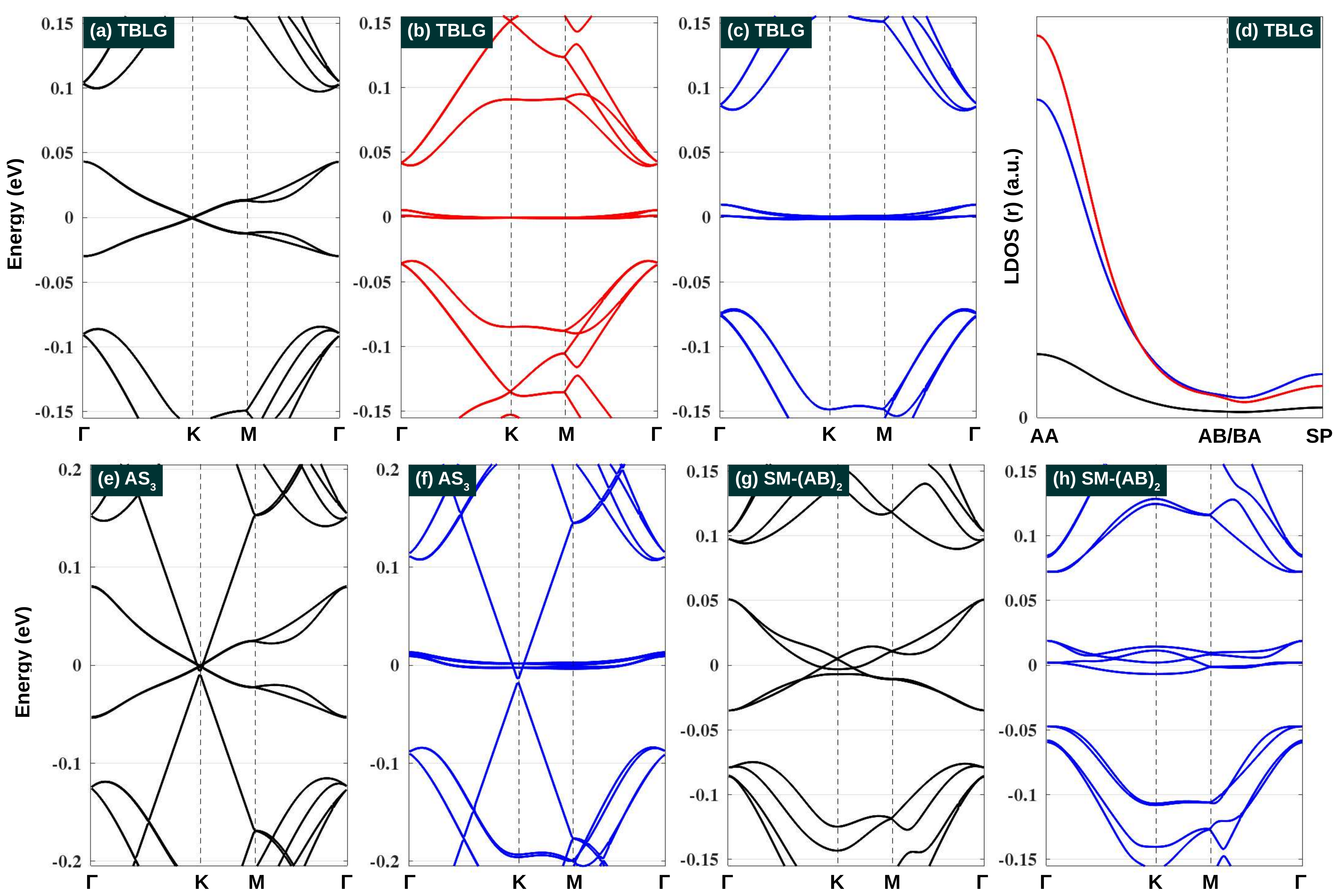}
		\caption{Electronic structures of TMGs modulated by strain and vertical pressure. TBLG at $\theta \simeq 1.30^\circ$ is considered in (a) for the pristine case, in (b) with a biaxial strain of 9$\%$, and in (c) with a vertical pressure $P \simeq 1.45$ GPa. The corresponding LDOS obtained at the strongest electronic localization point is presented in (d). Electronic bandstructures of AS$_3$ at $\theta \simeq 1.89^\circ$ with $P = 0$ and $\simeq$ 2.27 GPa is displayed in (e) and (f), respectively. The results of SM-(AB)$_2$ at $\theta \simeq 1.47^\circ$ with $P = 0$ and $\simeq$ 2.27 GPa are in (g) and (h), respectively.}
		\label{fig_sim14}
	\end{figure}
	First, it is worth reminding that based on continuum models \cite{rafi2011}, the first magic angle of TBLG can be estimated using the equation
	\begin{equation}\label{MagAng}
		\theta_{MA} \approx 3w/(2\pi t)
	\end{equation}
	where \textit{w} and \textit{t} stand for interlayer and intralayer tunneling strengths, respectively. Indeed, $\theta_{MA} \approx 1.1^\circ$ is obtained, in very good agreement with experiments \cite{cao2018a,yankowitz2019,gadelha2021,barbosa2022}, with $w = 0.109$ eV and $t = 2.7$ eV. Calculations using continuum models \cite{rafi2011,khalaf2019} have additionally predicted a discrete series of other magic angles $<1.1^\circ$. However, in agreement with the existing experiments (most clearly, see recent articles \cite{gadelha2021,barbosa2022}), atomistic calculations have demonstrated in Ref.~\cite{nguyen2021} and as discussed in section III that these small magic angles can not be practically obtained in reconstructed TBLGs. This is essentially due to the fact that while it gets maximum at $\theta \simeq 1.1^\circ$, the global electronic localization is progressively reduced, according to the increasing contribution of AB/BA stacking regions when reducing $\theta$ below $< 1.1^\circ$ as seen in Fig.\ref{fig_sim3} and more visibly in the left-panel of Fig.\ref{fig_sim13}. Interestingly, all these features are similarly observed in other TMGs as seen in two examples, i.e., AS$_3$ and SM-(AB)$_2$ systems, presented in Fig.\ref{fig_sim13}. Indeed, their electronic flat bands (i.e., magic angle) are exactly obtained when the real-space localization in AAA and AAB stacking regions, respectively, and their contribution to the global electronic properties of the system are concurrently maximized. Moreover, the commonness of such observation is further emphasized as it is also valid in TMGs when in-plane strains and/or vertical pressure are applied to tune the magic angle value (see Fig.\ref{fig_sim14} below, especially, the full illustration in Figs.\ref{fig_sim14}(a-d)). On the basis of those results, we thus conclude that the maximization of real-space electronic localization in regions where the AA stacking configuration is presented could be considered as the essential origin of the flat electronic bands observed in TMGs.
	
	In addition, the results in Fig.\ref{fig_sim13} are another illustration of feature discussed above that at their magic angles, the real-space electronic localization in the presence of Bernal bilayer graphene (i.e., the SM-(AB)$_2$ structure here) is significantly weaker than that in the systems of graphene monolayers only. This feature explained why less flat bands (accordingly, smaller localized DOS peak in energy) are obtained in magic-angle systems containing Bernal bilayer graphene, compared to other structures. 

	\begin{figure}[!ht]
	\centering
	\includegraphics[width = 0.59\textwidth]{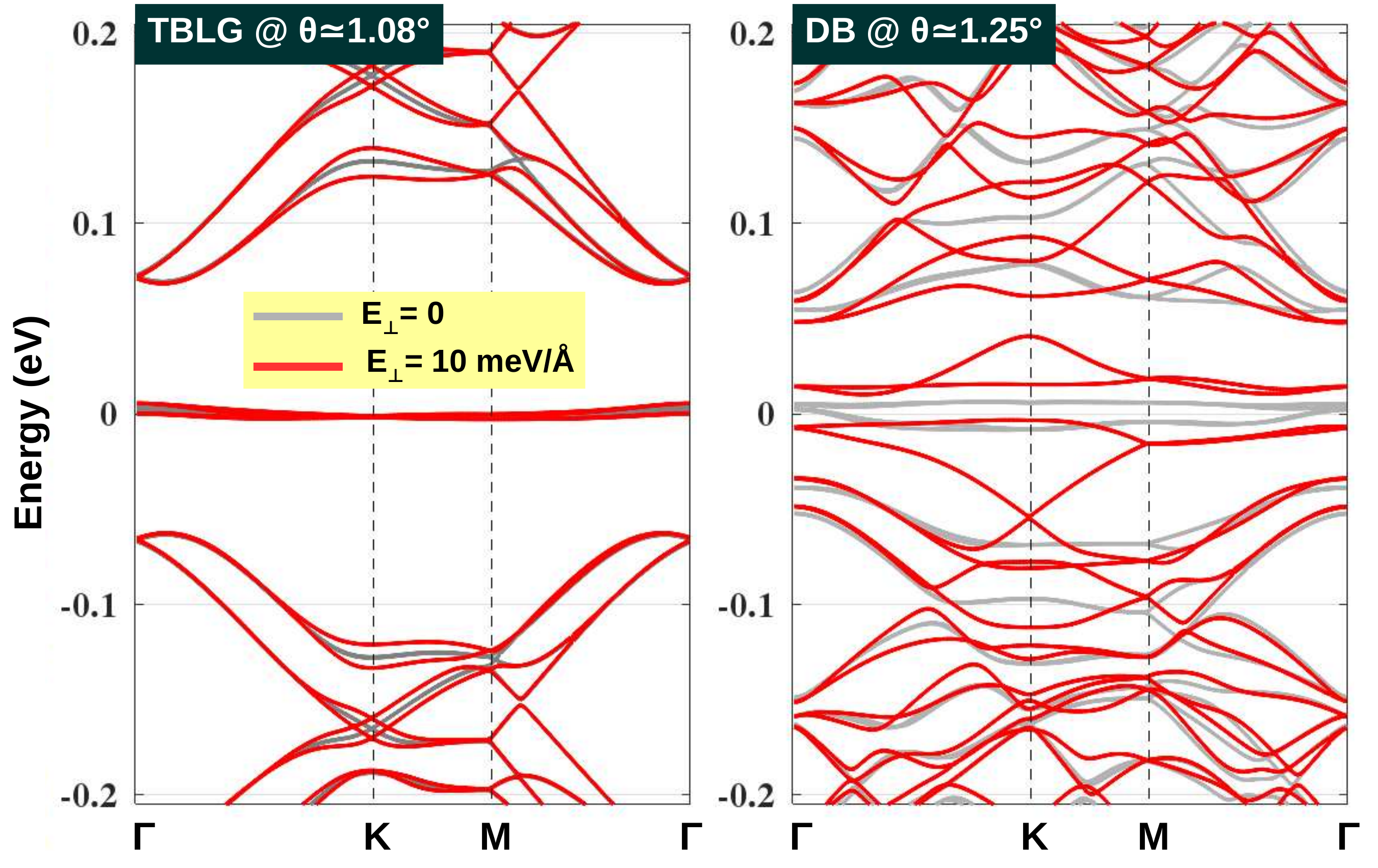}
	\caption{Effects of vertical electric field $E_\bot$ on the electronic bandstructure of TBLG at $\theta \simeq 1.08^\circ$ and DM-structure at $\theta \simeq 1.25^\circ$.}
	\label{fig_sim15}
    \end{figure}
    As mentioned, the dependence of magic-angle $\theta_{MA}$ on interlayer and intralayer tunneling strengths described in Eq.(\ref{MagAng}) suggests that $\theta_{MA}$ can be tuned by modifying $w$ and $t$, for instance, by applying in-plane strains and vertical pressure \cite{nguyen2015,huder2018,yankowitz2019,lin2020,liu2020s,wu2021,balint2021} as illustrated in Fig.\ref{fig_sim14}. In particular, the intralayer hopping energies are reduced by a tensile in-plane strain while the interlayer tunneling strength increases when a vertical pressure is applied \cite{pressure}, thus increasing $\theta_{MA}$ as shown. Even though they could be quantitatively different, these tunable possibilities are commonly obtained in all TMGs as illustrated in Fig.\ref{fig_sim14} for AS$_3$ and SM-(AB)$_2$ structures.
	
\begin{figure}[!ht]
		\centering
		\includegraphics[width = 0.98\textwidth]{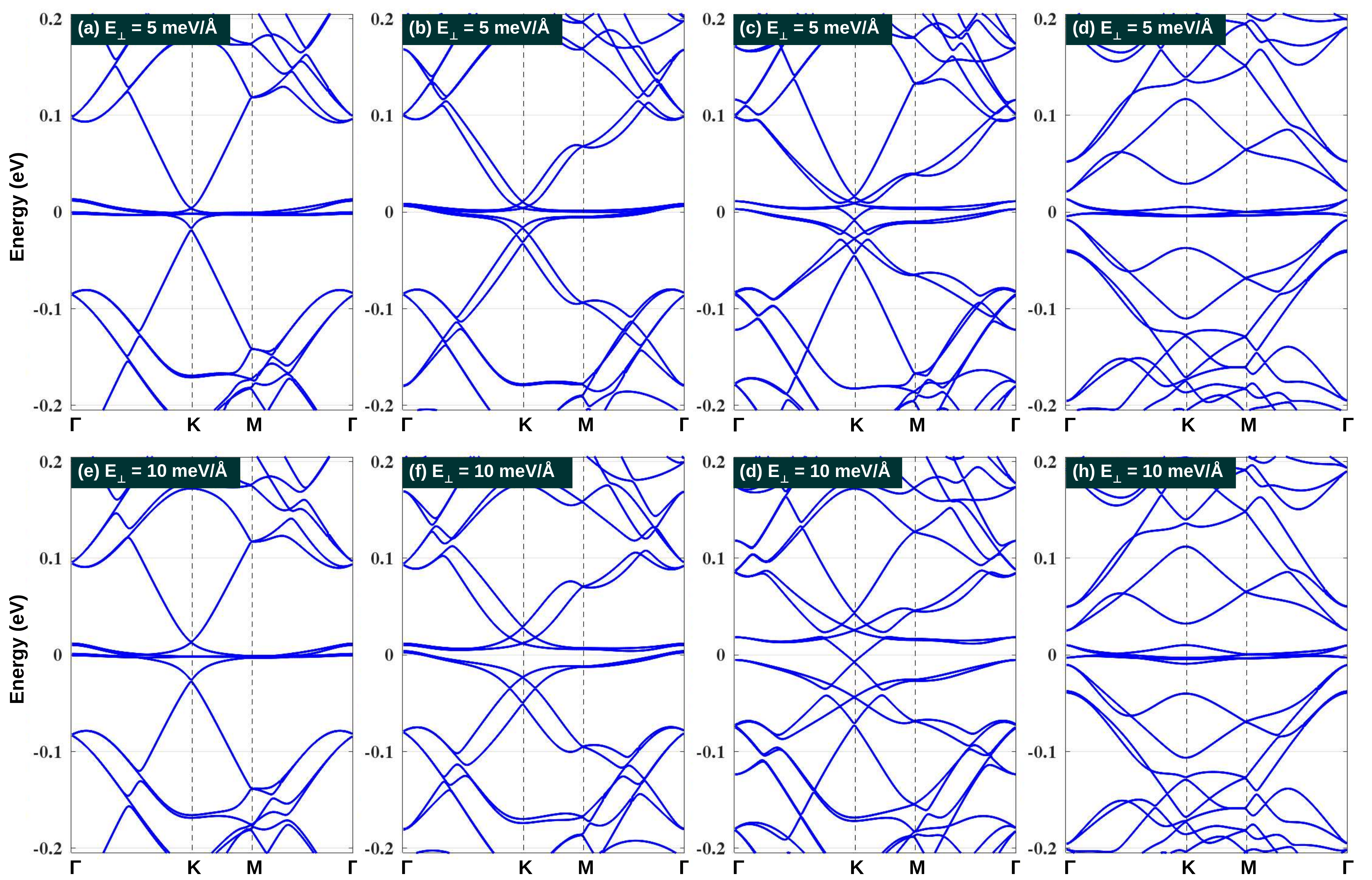}
		\caption{Electric field ($E_\bot$) effects on the electronic bandstructure of AS$_3$ at $\theta \simeq 1.54^\circ$ (a,e), AS$_4$ at $\theta \simeq 1.70^\circ$ (b,f), AS$_5$ at $\theta \simeq 1.79^\circ$ (c,g), and AS$_3$ considered in Fig.\ref{fig_sim9}a at $\theta \simeq 1.30^\circ$ (d,h).}
		\label{fig_sim16}
	\end{figure}

\begin{figure}[!b]
		\centering
		\includegraphics[width = 0.98\textwidth]{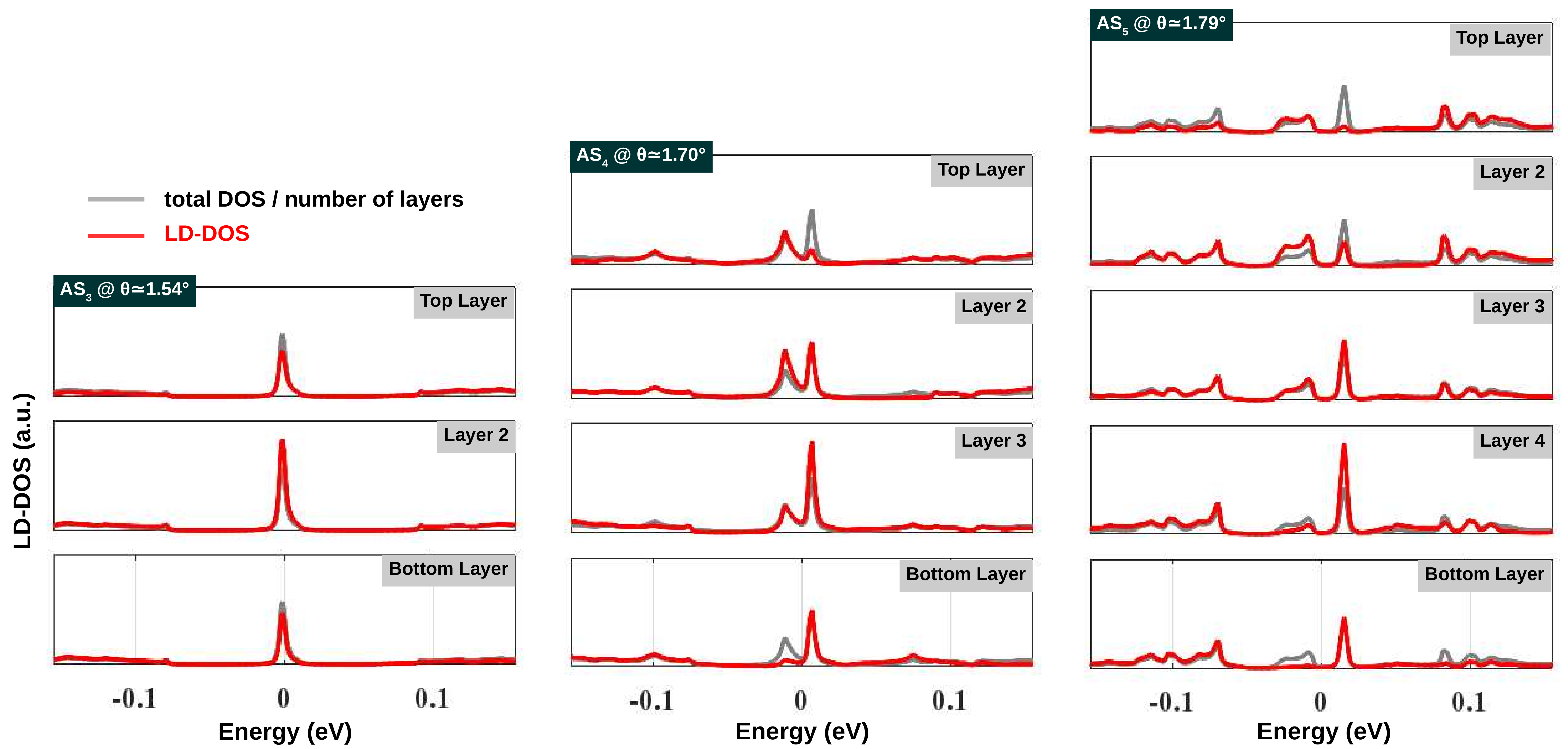}
		\caption{Layer decomposed DOS (LD-DOS) in systems considered in Figs.\ref{fig_sim16}(e-f) when an electric field $E_ \perp = 10$ meV/\AA \,\, is applied.}
		\label{fig_sim17}
	\end{figure}

	Another tunability has been actually explored \cite{bandwidth,liu2020,shen2020,chen2021e,rickhaus2021}, i.e., by a vertical electric field (\textit{E}-field) as illustrated in Figs.\ref{fig_sim15}-\ref{fig_sim17}. First, Fig.\ref{fig_sim15} presents typical pictures showing the \textit{E}-field effects in magic-angle twisted double bilayer graphene (i.e., DB-structure), compared to those in the TBLG structure. It shows that the \textit{E}-field generally presents stronger effects in DB-structure than in TBLG one. This could be simply understood as a direct consequence of the fact that \textit{E}-field can modulate the bandstructure and open a bandgap in Bernal stacking bilayer graphene, which is not observed for monolayer graphene. These results thus emphasize that vertical \textit{E}-field is a good degree of freedom to control the flat band and strongly correlated electronic phenomena in twisted systems containing Bernal stacking bilayer graphene, indeed as reported in Refs.~\cite{bandwidth,liu2020,shen2020,chen2021e,rickhaus2021}.
	
	Alternatively twisted multilayer systems of graphene monolayers (i.e., AS structures) is another example that has shown a good tunability by vertical \textit{E}-field \cite{zeyu2021,park2021,jeong2021}. In particular, when an \textit{E}-field is applied, a distinct feature observed in AS structures is that the flat bands can strongly hybridize with the Dirac bands, leading to the controllability of the bandwidth and interaction strength in the flat bands. In Fig.\ref{fig_sim16}, the electronic bandstructure of AS$_n$ (n=3,4,5) is presented for two different \textit{E}-fields. In addition to the mentioned hybridization, some other interesting effects are found. Indeed, while two flat bands hybridize with two Dirac bands, two other flat bands are still observed in the AS$_3$ case. For the AS$_4$ systems, four Dirac bands are observed at low energies (see Fig.\ref{fig_sim7}) and hence the hybridization takes place for all four flat bands with these four Dirac bands when a large \textit{E}-field is applied. Moreover, the hybridized flat bands can be separated in energy by the \textit{E}-field. Similar features occur for AS$_n$ systems with $n \geqslant 5$, except that besides four Dirac bands hybridizing with flat bands, the remaining Dirac bands are still observed near the Fermi level and a larger separation of hybridized flat bands is obtained.
	
	For the AS$_n$ structures investigated in Fig.\ref{fig_sim9} when a $\theta_{60}$ rotation is applied, the \textit{E}-field even presents different effects, compared to the conventional cases in Fig.\ref{fig_sim7}. In particular, instead of the band hybridization, flat bands in such systems are increasingly isolated from other dispersive bands when an \textit{E}-field is applied, as seen in Fig.\ref{fig_sim16}(d,h) for the AS$_3$ system of Fig.\ref{fig_sim9}(a). These results could be simply understood by its different structural properties, compared to the conventional AS$_3$ one of Fig.\ref{fig_sim7} as analyzed in Fig.\ref{fig_sim10}.
	
	Finally, to clarify more completely those observed electronic features of AS$_n$ systems, we analyze their layer decomposed DOS, when an \textit{E}-field is applied, in Fig.\ref{fig_sim17}. In the AS$_3$ case, even though the band hybridization occurs, changes in LD-DOS are relatively weak at the considered \textit{E}-fields. This could be because besides two hybridized flat bands, other two flat bands are still preserved near the Fermi level as mentioned above. Stronger \textit{E}-field effects on LD-DOS are found in the cases of a larger number of layers. In particular, the single localized DOS peak at zero \textit{E}-field is generally separated into two peaks when an \textit{E}-field is applied. Interestingly, those two peaks are shown to be likely separated in different layers. Indeed, the electron localized peak is mostly observed in the bottom layers whereas the hole peak in the top ones in the case of AS$_4$ and AS$_5$ systems in Fig.\ref{fig_sim17}. These results suggest possibilities to control the spatial dependence of electron localization and therefore of strongly correlated electronic phenomena in those twisted systems.
	
	\section{Summary}
	Using atomistic calculations, the electronic properties of twisted multilayer graphene systems are systematically studied. First, it is shown that both their structural and electronic properties exhibit a common feature that there is a unique magic angle at which the electronic localization in regions containing AA stacking configuration and its contribution to the global properties of the system are concurrently maximized. Flat electronic bands and strongly correlated electronic phenomena are therefore obtained at this critical case. In addition, the structural and electronic (particularly, electronic localization) properties are shown to exhibit a strong correlation. Consequently, the magic angle separates TMG systems into two classes, i.e., in small and large angle regimes, exhibiting distinct properties as clearly illustrated by the presented results obtained in TBLG \cite{nguyen2021}. Moreover, the strong correlation between the maximization of electronic localization in AA stacking region and the observation of flat bands presents an unknown physical view about the origin of magic angle. The electronic properties of magic-angle TMGs containing more than 2 graphene layers are shown to exhibit some other distinct properties, due to the presence of a large number of layers as well as various stacking configurations. In particular, the flat bands and dispersive ones at low energies can be concurrently obtained in some structures such as SM-(AB)$_n$ and alternatively twisted ones. The coexistence of localized-delocalized electronic states spatially separated are observed. At last, these TMGs present good possibilities for tuning electronic properties by external fields. Finally, the magic angle is also shown to be nicely tunable by the effects of strain and/or vertical pressure. Our study thereby provides a comprehensive overview as well as highlights the essential and outstanding electronic properties of twisted graphene systems that could be helpful for further developments in \textit{twistronics} research.

    
	\section*{Acknowledgments} V.-H.N. and J.-C.C. acknowledge financial support from the F\'ed\'eration Wallonie-Bruxelles through the ARC Grants (N$^{\circ}$ 16/21-077 and N$^{\circ}$ 21/26-116), from the European Union’s Horizon 2020 Research Project and Innovation Program — Graphene Flagship Core3 (N$^{\circ}$ 881603), from the Flag-Era JTC project “TATTOOS” (N$^{\circ}$ R.8010.19), from the EOS project “CONNECT” (N$^{\circ}$ 40007563) and from the Belgium F.R.S.-FNRS through the research projects (N$^{\circ}$ T.0051.18 and N$^{\circ}$ T.029.22F). Computational resources have been provided by the CISM supercomputing facilities of UCLouvain and the C\'ECI consortium funded by F.R.S.-FNRS of Belgium (N$^{\circ}$ 2.5020.11). T.X.H. acknowledges support from International Centre of Physics at the Institute of Physics, VAST, under grant number ICP.2022.05.

\bibliographystyle{unsrt}
\bibliography{mNguyen_biblio}

\end{document}